\documentclass[preprint,a4paper,11pt]{elsarticle}
\usepackage[T1]{fontenc}
\usepackage[utf8]{inputenc}
\usepackage[english]{babel}
\usepackage{amsmath,amsthm,enumerate}
\usepackage{amssymb}
\usepackage{amscd}
\usepackage{vmargin}
\usepackage{url}
\usepackage{stmaryrd}
\usepackage{tikz,pgf}
\usepackage{subfig}
\usepackage{cancel}
\usetikzlibrary{positioning}
\usepackage{subfig}
\usetikzlibrary{shapes.geometric}
\usepackage{supertabular}
\usepackage{comment}
  
\begin{document}
\begin{frontmatter}

\title{On the signed chromatic number of some classes of graphs}

\author[nice]{Julien Bensmail}
\author[isi]{Sandip Das}
\author[iem]{Soumen Nandi}
\author[labri,muni,liris]{\\Théo Pierron}
\author[iit]{Sagnik Sen}
\author[labri]{Éric Sopena}

\address[nice]{Universit\'e C\^ote d'Azur, CNRS, Inria, I3S, France}
\address[isi]{Indian Statistical Institute, Kolkata, India}
\address[iem]{Institute of Engineering \& Management, Kolkata, India}
\address[labri]{Univ. Bordeaux, Bordeaux INP, CNRS, LaBRI, UMR5800, F-33400 Talence, France}
\address[iit]{Indian Institute of Technology Dharwad, India}
\address[muni]{Faculty of Informatics, Masaryk University, Botanick\'a 68A, 602 00 Brno, Czech Republic}
\address[liris]{Université Lyon 1, LIRIS, UMR CNRS 5205, F-69621 Lyon, France}

\journal{...}

\begin{abstract} 
A signed graph $(G, \sigma)$ is a graph $G$ along with a function $\sigma: E(G) \to \{+,-\}$. A closed walk of a signed graph is positive (resp., negative) if it has an even (resp., odd) number of negative edges, counting repetitions. 
A homomorphism of a (simple) signed graph to another signed graph is a vertex-mapping that preserves adjacencies and signs of closed walks. The signed chromatic number of a signed graph $(G, \sigma)$ is the minimum number of vertices $|V(H)|$ of a signed graph $(H, \pi)$
 to which $(G, \sigma)$ admits a homomorphism. 

Homomorphisms of signed graphs have been attracting growing attention in the last decades, especially due to their strong connections to the theories of graph coloring and graph minors. 
These homomorphisms have been particularly studied through the scope of the signed chromatic number.
In this work, we 
provide new results and bounds on the signed chromatic number of several families of signed graphs (planar graphs, triangle-free planar graphs, $K_n$-minor-free graphs, and bounded-degree graphs). 

\medskip 

\end{abstract}
 
\end{frontmatter}

\newtheorem{theorem}{Theorem}[section]
\newtheorem{lemma}[theorem]{Lemma}
\newtheorem{conjecture}[theorem]{Conjecture}
\newtheorem{observation}[theorem]{Observation}
\newtheorem{claim}[theorem]{Claim}
\newtheorem{corollary}[theorem]{Corollary}
\newtheorem{proposition}[theorem]{Proposition}
\newtheorem{question}[theorem]{Question}

\newcommand{\qedclaim}{\hfill $\diamond$ \medskip}
\newenvironment{proofclaim}{\noindent{\em Proof of the claim.}}{\qedclaim}

\def\CHISP{\chi_{sp}}


\section{Introduction} \label{section:introduction}

Naserasr, Rollov\'{a} and Sopena introduced and initiated in~\cite{nrs2} the study of \textbf{homomorphisms of signed graphs}, based on the works of Zaslavsky~\cite{Zaslavsky2} and Guenin~\cite{guenin}. Over the passed few years, their work has generated increasing attention to the topic, see e.g.~\cite{DBLP:journals/jct/BeaudouFN17,DBLP:journals/dm/BrewsterFHN17, Accepted, NRS1, DBLP:journals/gc/NaserasrSS16, DBLP:journals/jgt/OchemPS17}. 
One reason behind this interest lies in the fact that homomorphisms of signed graphs stand as a natural way for generalizing a number of classical results and conjectures from graph theory, including, especially, ones related to graph minor theory (such as the Four-Color Theorem and Hadwiger's Conjecture). 
More generally, signed graphs are objects that arise in many contexts. Quite recently, for instance, Huang solved the Sensitivity Conjecture in~\cite{sensitivity_huang}, through the use, in particular, of signed graphs. His result was later improved by Laplante, Naserasr and Sunny in~\cite{sensitivity_reza}. These interesting works and results brought yet more attention to the topic.

In the recent years, works on homomorphisms of signed graphs have developed following two main branches. The first branch of research deals with attempts to generalize existing results and to solve standing conjectures  (regarding mainly undirected graphs). The second branch of research aims at understanding the very nature of signed graphs and their homomorphisms, thereby developing its own theory. Since our investigations in this paper are not related to all those concerns, there would be no point giving an exhaustive survey of the whole field. Instead, we focus on the definitions, notions, and previous investigations that connect to our work. Still, we need to cover a lot of material to make our motivations and investigations understandable. To ease the reading, we have consequently split this section into smaller subsections with different contents.

\subsection{Signed graphs and homomorphisms}\label{subsection:def}

Throughout this work, we restrict ourselves to graphs that are \textit{simple}, i.e., loopless graphs in which every two vertices are joined by at most one edge. For modified types of graphs, such as signed graphs, the notion of simplicity is understood with respect to their underlying graph.
Given a graph $G$, as per usual, $V(G)$ and $E(G)$ denote the set of vertices and the set of edges, respectively, of $G$.

A \textit{signed graph} $(G, \sigma)$ is a graph $G$ along with a function 
$\sigma: E(G) \rightarrow \{+,-\}$ called its \textit{signature}.
For every edge $e \in E(G)$, we call $\sigma(e)$ the \textit{sign} of $e$.
The edges of $(G, \sigma)$ in $\sigma^{-1}(+)$ are \textit{positive}, while the edges in
$\sigma^{-1}(-)$ are  \textit{negative}. 
In certain circumstances, it will be more convenient to deal with $(G, \sigma)$ in such a way that its set of negative edges is emphasized, in which case we will write $(G, \Sigma)$ instead, where $\Sigma = \sigma^{-1}(-)$ denotes the set of negative edges. Note that the notations $(G,\sigma)$ and $(G,\Sigma)$ are equivalent anyway, since $\sigma$ can be deduced from $\Sigma$, and \textit{vice versa}.

Signed graphs come with a particular \textit{switching operation} that can be performed on sets of vertices. For a vertex $v \in V(G)$ of a signed graph $(G, \sigma)$, \textit{switching} $v$ means changing the sign of all the edges incident to $v$. This definition extends to sets of vertices: for a set $S \subseteq V(G)$ of vertices of $(G, \sigma)$, \textit{switching} $S$ means changing the sign of the edges in the cut $(S, V(G) \setminus S)$.
For $S \subseteq V(G)$, we denote by $(G, \sigma^{(S)})$ the signed graph obtained from $(G, \sigma)$ when switching $S$. Two signed graphs $(G, \sigma_1)$ and $(G, \sigma_2)$ are  \textit{switching-equivalent} if $(G, \sigma_2)$ can be obtained from $(G, \sigma_1)$ by switching a set of vertices, which we write $(G, \sigma_1) \sim (G, \sigma_2)$. Note that $\sim$ is indeed an equivalence relation. 

An important notion in the study of signed graphs is the sign of its closed walks. Recall that, in a graph, a \textit{walk} is a path in which vertices and edges can be repeated. A \textit{closed walk} is a walk starting and ending at the same vertex. A closed walk $C$ of a signed graph is \textit{positive} if it has an even number of negative edges (counting with multiplicity), and \textit{negative} otherwise.
Observe that the sign of closed walks is invariant under the switching operation. 
In fact, the two notions are even more  related, as revealed by Zaslavsky's Lemma.

\begin{lemma}[Zaslavsky~\cite{Zaslavsky2}]\label{lem Zaslavsky}
Let $(G, \sigma_1)$ and $(G, \sigma_2)$ be two signed graphs having the same underlying graph $G$. 
Then $(G, \sigma_1) \sim (G, \sigma_2)$ if and only if the sign of every closed walk is the same 
in both $(G, \sigma_1)$ and $(G, \sigma_2)$.
\end{lemma}

Before moving on to all the definitions and notions related to signed graph homomorphisms, let us point out to the reader that the main difference between signed graphs and $2$-edge-colored graphs lies in the switching operation. Recall that a \textit{$2$-edge-colored graph} $(G,c)$ is a graph $G$ along with a function $c: E(G) \rightarrow \{1,2\}$ that assigns one of two possible colors to the edges, but with no switching operation. Thus, in some sense, $2$-edge-colored graphs stand as a static version (sign-wise) of signed graphs.
It was noticed in~\cite{switch_iso,nrs2,DBLP:journals/jgt/OchemPS17} that homomorphisms of $2$-edge-colored graphs are closely related to homomorphisms of signed graphs. For the sake of uniformity and convenience, we below refer to such homomorphisms as \textit{sign-preserving homomorphisms} of signed graphs. The study of such homomorphisms was initiated in~\cite{alon1998homomorphisms} independently from
the notion of homomorphisms of signed graphs. 

A \textit{sign-preserving homomorphism} (or \textit{sp-homomorphism} for short) of a signed graph
$(G, \sigma)$ to a signed graph $(H,\pi)$
is a vertex-mapping $f: V(G) \rightarrow V(H)$ that preserves adjacencies and signs of edges, i.e.,
for every $uv \in E(G)$ we have 
$f(u)f(v) \in E(H)$ 
and  $\sigma(uv) = \pi(f(u)f(v))$. 
When such an sp-homomorphism exists, we write $(G, \sigma) \xrightarrow{sp} (H, \pi)$.
The \textit{sign-preserving chromatic number} $\CHISP((G,\sigma))$ of a signed graph 
$(G,\sigma)$ 
is the minimum order $|V(H)|$ of a signed graph $(H, \pi)$ such that $(G,\sigma) \xrightarrow{sp} (H, \pi)$. 
For a family $\mathcal{F}$ of graphs, the sign-preserving chromatic number is generalized as 
$$\CHISP(\mathcal{F}) = \max\left\{\CHISP((G,\sigma)) : G \in \mathcal{F}\right\}.$$
A signed graph $(H, \pi)$ is said to be a \textit{sign-preserving bound} (or \textit{sp-bound} for short) of $\mathcal{F}$ if $(G, \sigma) \xrightarrow{sp} (H, \pi)$ for all $G \in \mathcal{F}$.  Furthermore, $(H, \pi)$ is \textit{minimal} if no proper subgraph of $(H, \pi)$ is an sp-bound of $\mathcal{F}$.

\medskip

We are now ready to define homomorphisms of signed graphs. It is worth mentioning that the upcoming definition is a restricted simpler version of a more general one~\cite{signed_hom_update}. 
A \textit{homomorphism} of a signed graph $(G, \sigma)$ to a signed graph $(H, \pi)$ 
is a vertex-mapping $f: V(G) \rightarrow V(H)$ that preserves adjacencies and  signs of  closed walks. 
We write $(G, \sigma) \rightarrow (H, \pi)$ whenever $(G, \sigma)$ admits a homomorphism to $(H, \pi)$.

 The next proposition highlights the underlying connection between sp-homomorphisms and homomorphisms of signed graphs. This proposition actually provides an alternative definition of homomorphisms of signed graph.

 \begin{proposition}[Naserasr, Sopena, Zaslavsky~\cite{signed_hom_update}]
 A mapping $f$ is a homomorphism of $(G, \sigma)$ to $(H, \pi)$ if and only if there exists
 $(G, \sigma') \sim (G, \sigma)$ such that $f$ is an sp-homomorphism of $(G, \sigma')$ to $(H, \pi)$. 
 \end{proposition}

Just as for sp-homomorphisms,
the \textit{signed chromatic number} $\chi_s((G,\sigma))$ of a signed graph 
$(G,\sigma)$ 
is the minimum order $|V(H)|$ of a signed graph $(H, \pi)$ 
such that $(G,\sigma) \rightarrow (H, \pi)$. 
For a family $\mathcal{F}$ of graphs, the signed chromatic number is given by 
$$\chi_s(\mathcal{F}) = \max\left\{\chi_s((G,\sigma)) : G \in \mathcal{F}\right\}.$$
Moreover, a \textit{bound} of $\mathcal{F}$ is a signed graph $(H, \pi)$   such that 
$(G, \sigma) \rightarrow (H, \pi)$ for all $G \in \mathcal{F}$. A bound of $\mathcal{F}$ is \textit{minimal} if none of its proper subgraphs is a bound of $\mathcal{F}$.

\subsection{Sign-preserving homomorphisms vs. homomorphisms of signed graphs}\label{subsection:connections}

One can observe that if $(G, \sigma)$ is a signed graph having positive edges (resp., negative edges)  only, 
then $(G, \sigma) \rightarrow (K_{\chi(G)}, \pi)$, where
$\chi(G)$ denotes the usual chromatic number of the graph $G$ and $(K_{\chi(G)}, \pi)$ is the signed complete graph of order $\chi(G)$ having positive edges (resp., negative edges) only,
 and thus $\CHISP((G, \sigma)) = \chi_s((G, \sigma)) = \chi(G)$. Hence, 
 the notions of sign-preserving chromatic number and signed chromatic number are indeed generalizations of the usual notion of chromatic number.

For undirected graphs, homomorphism bounds of minimum order are nothing but complete graphs. The study of sp-bounds and bounds for signed graphs is thus much richer from that point of view, as one of the most challenging aspects behind determining $\CHISP(\mathcal F)$ or $\chi_s(\mathcal F)$ for a given family $\mathcal F$ can actually be narrowed to finding (sp-)bounds of minimum order.

\medskip

One hint on the general connection between the sign-preserving chromatic number and the signed chromatic number is provided by the following results.

\begin{lemma}[Naserasr, Rollov\'{a}, Sopena~\cite{nrs2}]
Let $(G,\sigma)$ and $(H,\pi)$ be two signed graphs.
If $(G,\sigma) \rightarrow (H,\pi)$, then, for every $(H,\pi') \sim (H,\pi)$,
there exists $(G, \sigma') \sim (G, \sigma)$ such that 
$(G, \sigma') \xrightarrow{sp} (H, \pi')$. 
\end{lemma}

The connection between the sign-preserving chromatic number and the signed chromatic number was shown to be actually even deeper, through the concept of double switching graphs. Given a signed graph $(G, \sigma)$, 
the \textit{double switching graph $(\hat{G}, \hat{\sigma})$ of $(G, \sigma)$} is obtained from $(G, \sigma)$ by adding 
an \textit{anti-twin vertex} $\hat{v}$ for every vertex $v \in V(G)$, which means that for every $uv \in E(G)$, the graph $\hat{G}$ contains the edges $uv, u\hat{v}, \hat{u}v, \hat{u}\hat{v}$ and their signs satisfy
$\sigma(uv)=\hat{\sigma}(uv) = \hat{\sigma}(\hat{u}\hat{v}) \neq \hat{\sigma}(u\hat{v}) = \hat{\sigma}(\hat{u}v)$. One connection between $(G, \sigma)$ and $(\hat{G}, \hat{\sigma})$ is the following.

\begin{theorem}[Ochem, Pinlou, Sen~\cite{DBLP:journals/jgt/OchemPS17}]\label{th Ochem equiv}
For every two signed graphs $(G, \sigma)$ and $(H, \pi)$, we have $(G, \sigma) \rightarrow (H, \pi)$ if and only if 
$(G, \sigma) \xrightarrow{sp} (\hat{H}, \hat{\pi})$.
\end{theorem}

In particular, this result implies the following relations between the two chromatic numbers. 
\begin{proposition}[Naserasr, Rollov\'{a}, Sopena~\cite{nrs2}]\label{prop 2-ec signed}
For every signed graph $(G,\sigma)$, we have 
$\chi_s((G,\sigma)) \leq \CHISP((G,\sigma)) \leq 2 \chi_s((G,\sigma))$.
\end{proposition}

An even deeper connection, based on (sp-)isomorphisms 
of signed graphs, was established by Brewster and Graves~\cite{switch_iso}. 
A bijective (sp-)homomorphism whose inverse is also an (sp-)homomorphism is an \textit{(sp-)isomorphism}. 
Two signed graphs are \textit{(sp-)isomorphic} if there exists an (sp-)isomorphism between the two.

\begin{theorem}[Brewster, Graves~\cite{switch_iso}]\label{th rick equiv}
Two signed graphs $(G, \sigma)$ and $(H, \pi)$ are isomorphic if and only if 
$(\hat{G}, \hat{\sigma})$ and $(\hat{H}, \hat{\pi})$ are sp-isomorphic.
\end{theorem}


\subsection{Our contribution} \label{subsection:intro-contribution}

In this paper, we establish bounds and results related to the sign-preserving chromatic number and the signed chromatic number of various families of graphs. More precisely, we focus on planar graphs with given girth, $K_n$-minor-free graphs, and graphs with bounded maximum degree. Each of our results is proved in a dedicated section.

\subsubsection*{Planar graphs}

Recall that the \textit{girth} of a graph refers to the length of its shortest cycles.
We denote by $\mathcal{P}_g$ the family of planar graphs having girth at least $g$. Then, note that $\mathcal{P}_3$ is nothing but the whole family of planar graphs,
while $\mathcal{P}_4$ is the family of triangle-free planar graphs.

Towards establishing analogues of the Four-Color Theorem and of Gr\"{o}tzsch's Theorem for signed graphs, several works have been dedicated to studying the parameters $\CHISP(\mathcal{P}_g)$ and $\chi_s(\mathcal{P}_g)$. 
Note that it is worthwhile investigating such aspects, since,
for all values of $g \geq 3$, these two chromatic parameters are known to be finite, due to the existence of an (sp-)bound of $\mathcal{P}_g$ (see~\cite{DBLP:journals/jgt/OchemPS17}).

Let us now discuss the best known bounds on $\CHISP(\mathcal{P}_g)$ and $\chi_s(\mathcal{P}_g)$ for small values of~$g$.
Regarding the whole family $\mathcal{P}_3$ of planar graphs, it is known that $20 \leq \CHISP(\mathcal{P}_3) \leq 80$ and $10 \leq \chi_s(\mathcal{P}_3) \leq 40$ hold, as proved in~\cite{alon1998homomorphisms} and~\cite{DBLP:journals/jgt/OchemPS17}, respectively. 
In particular, it is worth mentioning  that 
if  $\chi_s(\mathcal{P}_3) = 10$, 
then there even exists a bound of  $\mathcal{P}_3$ of order~$10$. 
Ochem, Pinlou and Sen have actually shown in~\cite{DBLP:journals/jgt/OchemPS17} that if such a bound exists, it must be isomorphic to $(SP_9^+, \square^+)$, a signed graph we describe in upcoming Section~\ref{section:preliminaries}.
Due to Theorems \ref{th Ochem equiv} and~\ref{th rick equiv}, one may equivalently express this result in the following fashion. 

\begin{theorem}[Ochem, Pinlou, Sen~\cite{DBLP:journals/jgt/OchemPS17}]
\label{th ochem unique}
If there is a double switching sp-bound of 
$\mathcal{P}_3$ of order~$20$, 
then it is sp-isomorphic to $(\hat{SP}_9^+, \hat{\square}^+)$. 
\end{theorem}

Our first main result in this work (proved in Section~\ref{sec 1}) is that Theorem~\ref{th ochem unique} can be strengthened, in the sense that it also holds when dropping the double switching requirement from the statement. That is, we show that the only possible minimal sp-bound of $\mathcal{P}_3$ of order~$20$ has to be $(\hat{SP}_9^+, \hat{\square}^+)$. 

\begin{theorem}\label{th 2ec planar bound}
If there is a minimal sp-bound of $\mathcal{P}_3$ of order~$20$, then it is 
sp-isomorphic to 
$(\hat{SP}_9^+, \hat{\square}^+)$. 
\end{theorem}

It is worth mentioning that, supported by computer experimentations and theoretical evidences, it is conjectured that  $(SP_9^+, \square^+)$ is indeed 
a bound of $\mathcal{P}_3$ (see~\cite{classification}).

\subsubsection*{Triangle-free planar graphs}

For the family $\mathcal{P}_4$ of triangle-free planar graphs, it is known that $12 \leq \CHISP(\mathcal{P}_4) \leq 50$ and 
$6 \leq \chi_s(\mathcal{P}_4) \leq 25$ hold, as proved in~\cite{DBLP:journals/jgt/OchemPS17}. 
A natural intuition is that if $\chi_s(\mathcal{P}_3) = 10$ indeed held, then it would not be too surprising to have 
$\chi_s(\mathcal{P}_4) = 6$. In practice, however, making that step would not be that easy, as, in general, bounds are seemingly difficult to prove. From that point of view, it would be interesting to have an analogous version of 
Theorem~\ref{th 2ec planar bound} in this context. Our second main result in this paper (proved in Section~\ref{sec 2}) lies in that spirit, and reads as follows (where, again, the description of $(SP_5^+, \square^+)$, a signed graphs of order~$6$, is postponed to Section~\ref{section:preliminaries}).

\begin{theorem}\label{th triangle-free planar bound}
If there is a bound of $\mathcal{P}_4$ of order~$6$, then it is 
isomorphic to
$(SP_5^+, \square^+)$. 
\end{theorem}

\subsubsection*{$K_n$-minor-free graphs}

Let $\mathcal{F}_n$ denote the family of $K_n$-minor-free graphs. 
It is known that $\CHISP(\mathcal{F}_n) = 1, 4, 9$~\cite{alon1998homomorphisms} and $\chi_s(\mathcal{F}_n) = 1, 2, 5$ for $n = 2,3,4$, respectively~\cite{nrs2}. In~\cite{classification}, it was shown that if $\chi_s(\mathcal{P}_3) = 10$ (which would imply $\CHISP(\mathcal{P}_3) = 20$) held, then it would imply
$\chi_s(\mathcal{F}_5) = 10$ (and thus $\CHISP(\mathcal{F}_5) = 20$ as well).
However, prior to studying (sp-)bounds of $\mathcal{F}_n$, a first significant step could be to first investigate analogues of Hadwiger's Conjecture. To progress towards such analogues, it is first important to investigate what types of lower and upper bounds of $\CHISP(\mathcal{F}_n)$ and $\chi_s(\mathcal{F}_n)$ one can expect. In particular, are  $\CHISP(\mathcal{F}_n)$ and $\chi_s(\mathcal{F}_n)$ upper bounded at all? In this work, our third main result (proved in Section~\ref{sec 3}) is the following series of results towards those concerns.
 
\begin{theorem}\label{th Kn-minor-free}
The following inequalities hold:
\begin{enumerate}[(i)]
\item For all $n \geq 3$, $$ \chi_s(\mathcal{F}_n) \leq 5 {n-1 \choose 2} 2^{5 {n-1 \choose 2} -2} \text{ ~~and~~ } \CHISP(\mathcal{F}_n) \leq 5 {n-1 \choose 2} 2^{5 {n-1 \choose 2} -1}.$$

\item For all $n \geq 2$,
$$  \CHISP(\mathcal{F}_n) \geq 
\begin{cases}
\frac{2^{n+1} - 5}{3},  &\text{ when $n$ is even},\\
\frac{2^{n+1} - 4}{3},  &\text{ when $n$ is odd}.
\end{cases}
$$

\item For all $n \geq 2$,
$$  \chi_s(\mathcal{F}_n) \geq 
\begin{cases}
\frac{2^{n} - 1}{3},  &\text{ when $n$ is even},\\
\frac{2^{n} - 2}{3},  &\text{ when $n$ is odd}.
\end{cases}
$$ 
\end{enumerate}
\end{theorem}

\subsubsection*{Graphs with given maximum degree}

Let $\mathcal{G}^{\rm c}_{\Delta}$ denote the family of connected graphs with maximum degree at most $\Delta$. 
For large values of $\Delta$, it is possible to mimic an existing proof from~\cite{push_delta} (related to the pushable chromatic number of oriented graphs\footnote{Without entering too much into the details, the reader should be aware of a parallel line of research dedicated to the so-called \textit{oriented chromatic number} and \textit{pushable chromatic number} of oriented graphs, which are, roughly speaking, a counterpart of the sign-preserving chromatic number and the signed chromatic number of signed graphs in which edges are oriented instead of signed. Although the studies of the signed chromatic number and of the oriented chromatic number are sometimes quite comparable, there exist contexts in which they actually differ significantly.
For instance, there exist undirected graphs with oriented chromatic number arbitrarily larger than their signed chromatic number, as well as undirected graphs with signed chromatic number arbitrarily larger than their oriented chromatic number~\cite{bensmail_duffy_sen}.}) to show the following result.

\begin{theorem}\label{th delta-toka}
For all $\Delta \geq 29$, we have $$2^{\frac{\Delta}{2} - 1} \leq \chi_s(\mathcal{G}^{\rm c}_{\Delta}) \leq (\Delta-3) \cdot (\Delta-1) \cdot 2^{\Delta-1} +2.$$
\end{theorem}

Since this result can be established by following the exact same lines as the proof from~\cite{push_delta}, there would be no point giving a proof, and we instead refer the reader to that reference.

For smaller values of $\Delta$, one natural question is whether one can come up with the exact value of $\chi_s(\mathcal{G}^{\rm c}_{\Delta})$. It is worth mentioning that the oriented analogue of this very question remains unanswered, even for the smallest values of $\Delta$ (see \cite{push_delta}). In the case of signed graphs, we answer that question for the case $\Delta=3$, which stands as our fourth main result (proved in Section~\ref{sec 4}) in this paper.

\begin{theorem}\label{th signed cubic}
We have
$\chi_s(\mathcal{G}^{\rm c}_{3}) = 6$.
\end{theorem}

In fact, we will prove the following stronger result that implies Theorem~\ref{th signed cubic} as a corollary.  In the statement, recall that $(SP_5^+, \square^+)$ refers to a signed graph that will be defined in Section~\ref{section:preliminaries};
the only important thing to know, at this point, is that it has order~$6$.

\begin{theorem}\label{th signed cubic+}
Every signed subcubic graph with no connected component isomorphic to $(K_4, \emptyset)$ or $(K_4, E(K_4))$ admits a homomorphism to $(SP_5^+, \square^+)$.
\end{theorem}


\section{Definitions, terminology, and preliminary results on Paley graphs} \label{section:preliminaries}

Perhaps one of the most challenging aspects of studying (sp-)homomorphisms of signed graphs is to exhibit (sp-)bounds. In what follows, we introduce a few popular such bounds that appeared in the literature, which are related to so-called \textbf{Paley graphs}.

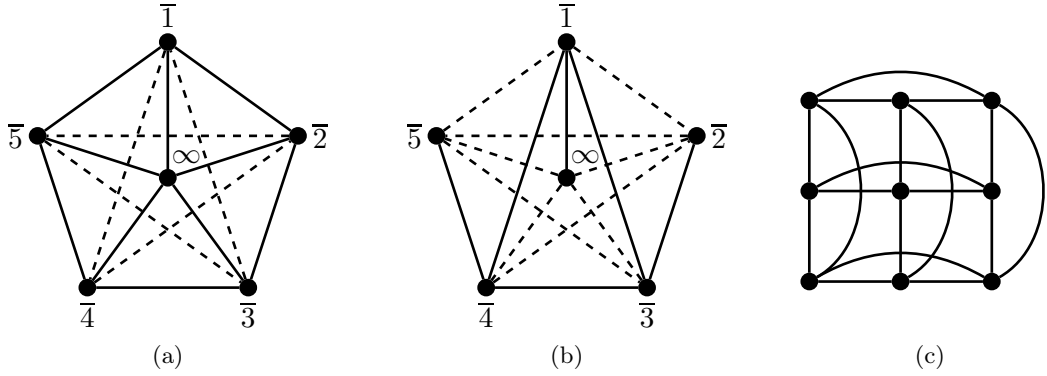
\begin{figure}[!ht]
 	\centering
 	\subfloat[\label{subfig:c1}]{
      	\begin{tikzpicture}[inner sep=0.7mm,scale=.9]
     		\node[draw, circle, fill, black, line width=1pt](v1) at (90:2)[label=above:{ $\overline{1}$}]{};
     		\node[draw, circle, fill, black,, line width=1pt](v2) at (18:2)[label=right:{ $\overline{2}$}]{};
     		\node[draw, circle, fill, black,, line width=1pt](v3) at (306:2)[label=below:{ $\overline{3}$}]{};
     		\node[draw, circle, fill, black, line width=1pt](v4) at (234:2)[label=below:{ $\overline{4}$}]{};
     		\node[draw, circle, fill, black, line width=1pt](v5) at (162:2)[label=left:{ $\overline{5}$}]{};
     		\node[draw, circle, fill, black, line width=1pt](vinf) at (0,0)[label={[xshift=-3pt,yshift=1pt] above right:{ $\infty$}}]{};
     		
     		\draw[-, line width=1pt,black] (v1) -- (v2) -- (v3) -- (v4) -- (v5) -- (v1); 
     		\draw[-, line width=1pt,black] (v1) -- (vinf); 
     		\draw[-, line width=1pt,black] (v2) -- (vinf); 
     		\draw[-, line width=1pt,black] (v3) -- (vinf); 
     		\draw[-, line width=1pt,black] (v4) -- (vinf); 
     		\draw[-, line width=1pt,black] (v5) -- (vinf); 

     		\draw[-, line width=1pt,black,dashed] (v1) -- (v4) -- (v2) -- (v5) -- (v3) -- (v1);
     	\end{tikzpicture}
    }
    \hspace{15pt}
 	\subfloat[\label{subfig:c1}]{
      	\begin{tikzpicture}[inner sep=0.7mm,scale=.9]
     		\node[draw, circle, fill, black, line width=1pt](v1) at (90:2)[label=above:{ $\overline{1}$}]{};
     		\node[draw, circle, fill, black,, line width=1pt](v2) at (18:2)[label=right:{ $\overline{2}$}]{};
     		\node[draw, circle, fill, black,, line width=1pt](v3) at (306:2)[label=below:{ $\overline{3}$}]{};
     		\node[draw, circle, fill, black, line width=1pt](v4) at (234:2)[label=below:{ $\overline{4}$}]{};
     		\node[draw, circle, fill, black, line width=1pt](v5) at (162:2)[label=left:{ $\overline{5}$}]{};
     		\node[draw, circle, fill, black, line width=1pt](vinf) at (0,0)[label={[xshift=-3pt,yshift=1pt] above right:{ $\infty$}}]{};
     		
     		\draw[-, line width=1pt,black] (v2) -- (v3) -- (v1) -- (vinf);
     		\draw[-, line width=1pt,black] (v3) -- (v4) -- (v1);
     		\draw[-, line width=1pt,black] (v4) -- (v5);

     		\draw[-, line width=1pt,black,dashed] (v4) -- (v2) -- (vinf) -- (v4);
     		\draw[-, line width=1pt,black,dashed] (v3) -- (vinf) -- (v5) -- (v1);
     		\draw[-, line width=1pt,black,dashed] (v1) -- (v2) -- (v5) -- (v3);
     	\end{tikzpicture}
    }
    \hspace{15pt}
 	\subfloat[\label{subfig:c1}]{
      	\begin{tikzpicture}[inner sep=0.7mm,scale=0.6]
     		\node[draw, circle, fill, black, line width=1pt](v1) at (0,0){};
     		\node[draw, circle, fill, black, line width=1pt](v2) at (0,2){};
     		\node[draw, circle, fill, black, line width=1pt](v3) at (0,4){};
     		
     		\node[draw, circle, fill, black, line width=1pt](v4) at (2,0){};
     		\node[draw, circle, fill, black, line width=1pt](v5) at (2,2){};
     		\node[draw, circle, fill, black, line width=1pt](v6) at (2,4){};
     		
     		\node[draw, circle, fill, black, line width=1pt](v7) at (4,0){};
     		\node[draw, circle, fill, black, line width=1pt](v8) at (4,2){};
     		\node[draw, circle, fill, black, line width=1pt](v9) at (4,4){};
     		
     		\draw[-, line width=1pt,black] (v1) -- (v2) -- (v3);
     		\draw[-, line width=1pt,black] (v4) -- (v5) -- (v6);
     		\draw[-, line width=1pt,black] (v7) -- (v8) -- (v9);
     		
     		\draw[-, line width=1pt,black] (v1) -- (v4) -- (v7);
     		\draw[-, line width=1pt,black] (v2) -- (v5) -- (v8);
     		\draw[-, line width=1pt,black] (v3) -- (v6) -- (v9);
     		
     		\draw[-, line width=1pt,black] (v1) to[out=30,in=150] (v7);
     		\draw[-, line width=1pt,black] (v2) to[out=30,in=150] (v8);
     		\draw[-, line width=1pt,black] (v3) to[out=30,in=150] (v9);
     	
     		\draw[-, line width=1pt,black] (v1) to[out=30,in=-30] (v3);
     		\draw[-, line width=1pt,black] (v4) to[out=30,in=-30] (v6);
     		\draw[-, line width=1pt,black] (v7) to[out=30,in=-30] (v9);
     		\node at (2,-1) {};
     	\end{tikzpicture}
    }

\caption{The signed graph $(SP_5^+, \square^+)$ (a),
the signed graph $(SP_5^+, \boxdot^+)$ obtained by switching the vertices $\infty$ and $\overline{1}$ of $(SP_5^+, \square^+)$ (b), and the signed graph $(SP_9, \square)$ (c).
In (a), (b) and (c), solid edges are positive edges.
In (a) and (b), dashed edges are negative edges.
In (c), non-edges are negative edges.}\label{fig Paley}
\end{figure}

Let $q \equiv 1 \bmod 4$ be a prime power, and $\mathbb{F}_q$ be the finite field of order $q$. 
The \textit{signed Paley graph} $(SP_q, \square)$ of order $q$ is the signed graph with 
 set of vertices $V(SP_q) = \mathbb{F}_q$,
 set of positive edges $\square^{-1}(+) = \{uv: u-v \text{ is a square in } \mathbb{F}_q\}$, and 
 set of negative edges $\square^{-1}(-) = \{uv: u-v \text{ is not a square in } \mathbb{F}_q\}$. 
The \textit{signed Paley plus graph} $(SP_q^+, \square^+)$ of order $q+1$ is the signed graph obtained from $(SP_q, \square)$
by adding a vertex $\infty$ and making it adjacent to every other vertex 
through a positive edge. 
To avoid ambiguities, we will refer to a vertex $i \neq \infty$ of $SP_q$ or $SP_q^+$ by writing $\overline{i}$.
See Figure~\ref{fig Paley} for an illustration.

Signed Paley graphs, signed Paley plus graphs, and their respective double switching graphs are, in the literature, regularly used as (sp-)bounds. One reason for that is that these graphs have a very symmetric structure, resulting in  properties that are very useful when it comes to designing homomorphisms.
Such useful properties deal, in particular, with some particular notions of transitivity.
More precisely, a signed graph $(G, \sigma)$ is \textit{sign-preserving vertex-transitive} (or 
\textit{sp-vertex-transitive} for short) if, for every two vertices $u, v \in V(G)$, there exists an sp-isomorphism $f$ of $(G, \sigma)$ to itself such that $f(u) = v$. Furthermore, $(G, \sigma)$ is 
\textit{sign-preserving edge-transitive} (or 
\textit{sp-edge-transitive} for short) if, for every two edges $uv, u'v' \in E(G)$ with the same sign, there exists an sp-isomorphism $f$ of $(G, \sigma)$ to itself such that $f(u) = u'$ and $f(v) = v'$. Similarly, $(G, \sigma)$ is \textit{vertex-transitive} 
if, for every two vertices $u, v \in V(G)$, there exists an isomorphism $f$ of $(G, \sigma)$ to itself 
such that $f(u) = v$; while $(G, \sigma)$ is 
\textit{edge-transitive} if, for every two edges $uv, u'v' \in E(G)$, there exists an isomorphism $f$ of $(G, \sigma)$ to itself 
such that $f(u) = u'$ and $f(v) = v'$. 

\begin{proposition}[Ochem, Pinlou, Sen~\cite{DBLP:journals/jgt/OchemPS17}]
Let $q \equiv 1 \bmod 4$ be a prime power. Then:
\begin{enumerate}[(i)]
\item $(SP_q, \square)$ is sp-vertex-transitive and sp-edge-transitive;

\item $(SP_q^+, \square^+)$ is vertex-transitive and edge-transitive. 
\end{enumerate}
\end{proposition} 

Given a  positive edge $uv$ of a signed graph $(G, \sigma)$, we call $u$ a \textit{positive neighbor} of~$v$. Analogously, $u$ is a \textit{negative neighbor} of $v$ if $uv$ is a negative edge.
We denote by $N(v)$, $N^+(v)$ and $N^-(v)$ the sets of neighbors, 
positive neighbors, and negative neighbors, respectively, of $v$ in $(G, \sigma)$.
Analogously, we define the \textit{degree} $d(v)$, \textit{positive degree} $d^+(v)$, and \textit{negative degree} $d^-(v)$ of $v$ 
as  $|N(v)|$, $|N^{+}(v)|$  and $|N^{-}(v)|$, respectively.  Assuming $u$ and $v$ are two distinct vertices having a common neighbor $w$, we say that $u$ and $v$ \textit{agree on $w$} if $w \in N^\alpha(u) \cap N^\alpha(v)$ for some $\alpha \in \{-,+\}$. Conversely, we say that $u$ and $v$ \textit{disagree on $w$} if they do not agree on $w$.

Let $\vec{v} = (v_1, \dots, v_k)$ be a $k$-tuple of distinct vertices of $(G, \sigma)$ and 
let $\vec{\alpha} = (\alpha_1, \dots, \alpha_k) \in \{+,-\}^k$ be a 
$k$-vector with each of its elements being $+$ or $-$. 
We define the \textit{$\vec{\alpha}$-neighborhood}
of $\vec{v}$ as
$$N^{\vec{\alpha}}(\vec{v}) = \cap_{i=1}^k  N^{\alpha_i}(v_i).$$
Moreover, we say that $(G, \sigma)$ \textit{has property $P_{k,\ell}$}  
if, for every $k$-tuple $\vec{v}$ and every $k$-vector $\vec{\alpha}$, we have  
$|N^{\vec{\alpha}}(\vec{v})| \geq \ell$. 
We also define the \textit{negation} $-\vec{\alpha}$ of a $k$-vector $\vec{\alpha} = (\alpha_1, \dots, \alpha_k)$ as
$-\vec{\alpha} = (-\alpha_1, \dots, -\alpha_k)$ where $-\alpha_i = -$ if $\alpha_i = +$, and
$-\alpha_i = +$ otherwise. 
The \textit{switched $\vec{\alpha}$-neighborhood} 
of $\vec{v}$ is then
$$\hat{N}^{\vec{\alpha}}(\vec{v}) = N^{\vec{\alpha}}(\vec{v})  \cup N^{-\vec{\alpha}}(\vec{v}).$$
Lastly, we say that $(G, \sigma)$ \textit{has 
property $\hat{P}_{k,\ell}$}  
if, for every $k$-tuple $\vec{v}$ and every $k$-vector $\vec{\alpha}$, we have  
$|\hat{N}^{\vec{\alpha}}(\vec{v})| \geq \ell$. 
Notice that this property is invariant under the switching operation. 

It turns out that signed Paley graphs and signed Paley plus graphs also have the following interesting properties, which are very convenient ones for designing homomorphisms.

\begin{proposition}[Ochem, Pinlou, Sen~\cite{DBLP:journals/jgt/OchemPS17}]\label{prop property pnk}
Let $q \equiv 1 \bmod 4$ be a prime power. Then:
\begin{enumerate}[(i)]
\item $(SP_q, \square)$ has property $P_{1,\frac{q-1}{2}}$ and $P_{2,\frac{q-5}{4}}$;

\item $(SP_q^+, \square^+)$ 
has property $\hat{P}_{1,q}$, $\hat{P}_{2,\frac{q-1}{2}}$ and $\hat{P}_{3,\frac{q-5}{4}}$.
\end{enumerate}
\end{proposition}

\section{Proof of Theorem~\ref{th 2ec planar bound}}\label{sec 1}

Let $(T, \lambda)$ be a minimal sp-bound of $\mathcal{P}_3$ of order~$20$, assuming that such a signed graph exists. In this section, our goal is to show that $(T, \lambda)$ must be sp-isomorphic to $(\hat{SP}_9^+, \hat{\square}^+)$. To this end, we first use the following lemma to show that $\Delta(T)\in\{18,19\}$. We then deal with each of the two possible values of $\Delta(T)$ separately.


\begin{lemma}\label{lem nbrhood}
For every vertex $v$ of $(T, \lambda)$,  we have $d^{+}(v),d^-(v) \geq 9$. 
Moreover, if $d^{\alpha}(v) = 9$ for an $\alpha \in \{+,-\}$, 
then the induced subgraph $(T, \lambda)[N^{\alpha}(v)]$ is sp-isomorphic 
to $(SP_9, \square)$.
\end{lemma}

\begin{proof}
It is known (see~\cite{montejano}) that, for the family $\mathcal{O}_3$ of outerplanar graphs, we have
$\CHISP(\mathcal{O}_3) = 9$ and the only sp-bound 
of $\mathcal{O}_3$ of order $9$  is $(SP_9, \square)$.
Thus,  there exists an outerplanar signed graph $(O, \varphi)$ with  
$\CHISP((O, \varphi)) = 9$, and such that the only signed graph of order $9$  to which $(O, \varphi)$
admits an sp-homomorphism is $(SP_9, \square)$. 
Also, it is known from~\cite{alon1998homomorphisms} that there exists a planar signed graph $(P, \pi)$ with $\CHISP((P, \pi)) = 20$. 

Let us now consider the planar signed graph $(P', \pi')$ obtained as follows: 
start from $(P,\pi)$, and,
for every $v \in V(P)$, add a copy of $(O, \varphi)$ to the $+$-neighborhood of $v$ 
and another copy to the $-$-neighborhood of $v$ (see Figure~\ref{fig copiesv}).
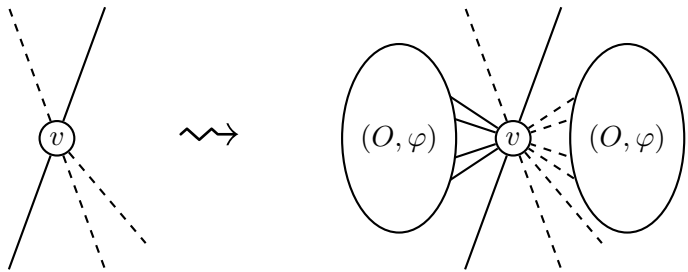
\begin{figure}[!ht]
\centering
\begin{tikzpicture}[thick]
\node[circle,draw,inner sep=2pt] (v) at (0,0) {$v$};
\node (v1) at (70:2) {};
\node (v2) at (110:2) {};
\node (v3) at (290:2) {};
\node (v4) at (250:2) {};
\node (v5) at (310:2) {};
\draw (v) -- (v1);
\draw[dashed] (v) -- (v2);
\draw[dashed] (v) -- (v3);
\draw (v) -- (v4);
\draw[dashed] (v) -- (v5);
\node at (2,0) {\Huge $\rightsquigarrow$};
\tikzset{xshift=6cm}
\node[circle,draw,inner sep=2pt] (v) at (0,0) {$v$};
\node (v1) at (70:2) {};
\node (v2) at (110:2) {};
\node (v3) at (290:2) {};
\node (v4) at (250:2) {};
\node (v5) at (310:2) {};
\draw (v) -- (v1);
\draw[dashed] (v) -- (v2);
\draw[dashed] (v) -- (v3);
\draw (v) -- (v4);
\draw[dashed] (v) -- (v5);

\draw(-1.5,1) -- (v);
\draw(-1.5,.5) -- (v);
\draw(-1.5,-.5) -- (v);
\draw(-1.5,-1) -- (v);
\draw[fill=white] (-1.5,0) ellipse (.75cm and 1.25cm);
\node[fill=white] at (-1.5,0) {$(O,\varphi)$};

\draw[dashed](1.5,1) -- (v);
\draw[dashed](1.5,.5) -- (v);
\draw[dashed](1.5,-.5) -- (v);
\draw[dashed](1.5,-1) -- (v);
\draw[fill=white] (1.5,0) ellipse (.75cm and 1.25cm);
\node[fill=white] at (1.5,0) {$(O,\varphi)$};

\end{tikzpicture}
\caption{Construction of $(P',\pi')$ in Lemma~\ref{lem nbrhood}. Solid edges are positive edges, dashed ones are negative.}
\label{fig copiesv}
\end{figure}

Observe that the so-obtained signed graph $(P', \pi')$ is planar. 
Also, according to our assumption,  $(P', \pi') \xrightarrow{sp} (T, \lambda)$. 
Therefore, for each $\alpha \in \{+, -\}$, we obtain 
\[d_T^{\alpha}(v)\geqslant \CHISP((P'[N^\alpha(v)],\pi'))=\CHISP((O,\varphi))=9.\] 
The last part of the statement follows from the fact that $(SP_9,\square)$ is
the only signed graph of order $9$ to which $(O, \varphi)$
admits an sp-homomorphism. 
\end{proof}

From the previous result, we deduce that the maximum degree $\Delta(T)$ of $T$ is $18$ or $19$. We first consider the case $\Delta(T)=18$, and show that $(T,\lambda)$ is sp-isomorphic to $(\hat{SP}_9^+,\square^+)$. Note that, by Theorem~\ref{th ochem unique}, we just have to show that $(T, \lambda)$ is a double switching graph.

\begin{lemma}
If $\Delta(T)=18$, then $(T, \lambda)$ is a double switching  graph.
\end{lemma}

\begin{proof}
Observe that Lemma~\ref{lem nbrhood} ensures that $\delta(T)=18$. Therefore, if $\Delta(T) = 18$, then $T$ is $18$-regular and thus has an anti-matching. 
Let now $v$ and $v'$ be two non-adjacent vertices of $T$, and assume that $v$ and $v'$ are not anti-twins. Then there exists a vertex $w\in N^{\alpha}(v) \cap N^{\alpha}(v')$ for some $\alpha\in\{+,-\}$. Observe now that $N^{\alpha}(w)$ contains both $v$ and $v'$, and hence induces a non-complete graph. This is a contradiction with Lemma~\ref{lem nbrhood} since $N^{\alpha}(w)$ induces $(SP_9,\square)$ whose underlying graph is complete. 
Therefore, every pair of non-adjacent vertices of $(T,\lambda)$ are anti-twins, implying that $(T,\lambda)$ is a double switching graph.
\end{proof}

This concludes the proof of Theorem~\ref{th 2ec planar bound} in the case $\Delta(T)=18$. The rest of this section is devoted to the case $\Delta(T)=19$, in which we aim for a contradiction. To obtain this contradiction, we investigate how the neighborhoods of adjacent vertices interact in $(T,\lambda)$.

\begin{lemma}\label{lem common nbrhood}
For every edge $uv$ of $(T, \lambda)$ and  $\alpha, \beta \in \{+, -\}$, 
we have $|N^{\alpha}(u) \cap N^{\beta}(v)| \geq 4$. 
\end{lemma}

\begin{proof}
As mentioned earlier, there exist planar signed graphs $(P, \pi)$ with $\CHISP((P, \pi)) = 20$,
and, because $(T, \lambda)$ is minimal,
that have the following property: 
 for every edge $uv \in E(T)$ and for 
 any sp-homomorphism $f: (P,\pi) \xrightarrow{sp} (T, \lambda)$,
there exists an edge $xy \in E(P)$ such that $f(x) = u$ and $f(y) = v$. 

Let $(P_5, M)$ denote the signed path on five edges whose three negative edges induce a maximum matching $M$. Observe that $\CHISP((P_5, M)) = 4$. 

Let us now consider the planar signed graph $(P', \pi')$ obtained as follows (see Figure~\ref{fig copiese}): 
start from $(P, \pi)$, and,
for every $xy \in E(P)$ and all $(\alpha, \beta) \in \{+,-\}^{2}$, include a copy of $(P_5, M)$
inside $N^{\alpha}(x) \cap N^{\beta}(y)$. 
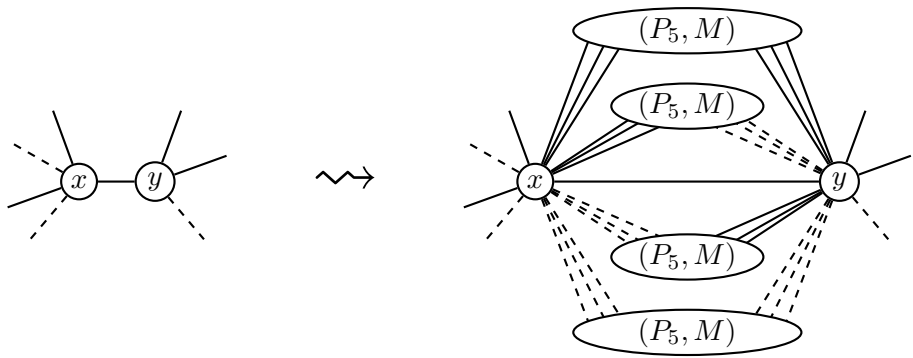
\begin{figure}[!ht]
\centering
\begin{tikzpicture}[thick]
\node[circle,draw,inner sep=2pt] (v) at (0,0) {$x$};
\draw (v) -- (110:1);
\draw[dashed] (v) -- (150:1);
\draw (v) -- (200:1);
\draw[dashed] (v) -- (230:1);
\tikzset{xshift=1cm}
\node[circle,draw,inner sep=2pt] (y) at (0,0) {$y$};
\draw (y) -- (70:1);
\draw (y) -- (20:1);
\draw[dashed] (y) -- (-50:1);
\draw (v) -- (y);
\node at (2.5,0) {\Huge $\rightsquigarrow$};
\tikzset{xshift=5cm}
\node[circle,draw,inner sep=2pt] (v) at (0,0) {$x$};
\draw (v) -- (110:1);
\draw[dashed] (v) -- (150:1);
\draw (v) -- (200:1);
\draw[dashed] (v) -- (230:1);
\draw (v) -- (1.25,2);
\draw (v) -- (.75,2);
\draw (v) -- (1,2);
\draw[dashed] (v) -- (1.25,-2);
\draw[dashed] (v) -- (.75,-2);
\draw[dashed] (v) -- (1,-2);
\draw (v) -- (1.5,1);
\draw (v) -- (1.75,1);
\draw (v) -- (2.25,1);
\draw[dashed] (v) -- (1.5,-1);
\draw[dashed] (v) -- (1.75,-1);
\draw[dashed] (v) -- (2.25,-1);

\tikzset{xshift=4cm}
\node[circle,draw,inner sep=2pt] (y) at (0,0) {$y$};
\draw (y) -- (70:1);
\draw (y) -- (20:1);
\draw[dashed] (y) -- (-50:1);
\draw (v) -- (y);
\draw (y) -- (-1.25,2);
\draw (y) -- (-.75,2);
\draw (y) -- (-1,2);
\draw[dashed] (y) -- (-1.25,-2);
\draw[dashed] (y) -- (-.75,-2);
\draw[dashed] (y) -- (-1,-2);
\draw (y) -- (-1.5,-1);
\draw (y) -- (-1.75,-1);
\draw (y) -- (-2.25,-1);
\draw[dashed] (y) -- (-1.5,1);
\draw[dashed] (y) -- (-1.75,1);
\draw[dashed] (y) -- (-2.25,1);

\draw[fill=white] (-2,2) ellipse (1.5cm and .3cm);
\draw[fill=white] (-2,1) ellipse (1cm and .3cm);
\draw[fill=white] (-2,-1) ellipse (1cm and .3cm);
\draw[fill=white] (-2,-2) ellipse (1.5cm and .3cm);
\node at (-2,2) {$(P_5,M)$};
\node at (-2,1) {$(P_5,M)$};
\node at (-2,-1) {$(P_5,M)$};
\node at (-2,-2) {$(P_5,M)$};

\end{tikzpicture}
\caption{Construction of $(P',\pi')$ in Lemma~\ref{lem common nbrhood}. Solid edges are positive edges, dashed ones are negative.}
\label{fig copiese}
\end{figure}
Observe that the so-obtained signed graph $(P', \pi')$ is planar. 
Furthermore, according to our assumption,  $(P', \pi') \xrightarrow{sp} (T, \lambda)$. 
Therefore, for every $uv \in E(T)$ and for all  $\alpha, \beta \in \{+, -\}$, 
every copy of  $(P_5, M)$ must admit a homomorphism to the  subgraph of $(T, \lambda)$ induced by 
$N^{\alpha}(u) \cap N^{\beta}(v)$. Hence, the fact that $\CHISP((P_5, M)) = 4$ implies that
$|N^{\alpha}(u) \cap N^{\beta}(v)| \geq 4$.
\end{proof}

In view of Lemmas~\ref{lem nbrhood} and~\ref{lem common nbrhood}, every intersection $N^\alpha(u)\cap N^\beta(v)$ induces a complete subgraph of $(SP_9,\square)$ of order at least~$4$. Before completing the proof, we investigate the possible signatures of the $K_4$'s that are subgraphs of $(SP_9,\square)$, and state some of their properties. Since these properties are easy to verify due to the vertex-transitivity and edge-transitivity of $(SP_9, \square)$, some formal proofs are omitted.

Let $(K_4, M^-)$ be the signed graph having the complete graph $K_4$ as its underlying graph and 
a perfect matching as its set of negative edges. Similarly, let 
$(K_4, M^+)$ be the signed graph having $K_4$ as its underlying graph and 
the edges of a  $4$-cycle (that is, the complement of a perfect matching) as its set of negative edges.

\begin{observation}\label{obs neighborhood of SP9}
For every vertex $v$ of $(SP_9, \square)$, the set $N^{+}(v)$ induces $(K_4, M^{+})$ in $(SP_9, \square)$, while the set $N^{-}(v)$ induces $(K_4, M^{-})$. 
\end{observation}

\begin{observation}\label{obs neighborhood of SP9 reverse}
For every induced $(K_4, M^{+})$ (resp., $(K_4, M^{-})$) of $(SP_9, \square)$, 
there exists a $v \in V(SP_9)$ such that 
 $N^{+}(v)$ (resp.,  $N^{-}(v)$) induces that  $(K_4, M^{+})$ (resp.,   $(K_4, M^{-})$). 
\end{observation}

We are now ready to derive the desired contradiction in the case $\Delta(T)=19$. We first need to show that there are two vertices $u$ and $v$ of ``large'' degree.

\begin{lemma}\label{lem uv property}
For some $\{\alpha, \overline{\alpha}\} = \{+,-\}$, there exists an $\alpha$-edge $uv$ of $(T, \lambda)$ such that 
$d^{\alpha}(u) = d^{\overline{\alpha}}(v) = 10$ and $d^\alpha(v)=9$.
\end{lemma}

\begin{proof}
Since $\Delta(T)=19$, there is a vertex $v\in V(T)$ with $d(v) = 19$. By Lemma~\ref{lem nbrhood}, we have
$d^{\alpha}(v) = 9$ and $d^{\overline{\alpha}}(v) = 10$ for some $\{\alpha, \overline{\alpha}\} = \{+,-\}$. 
By Lemma~\ref{lem common nbrhood}, each vertex in $N^{\overline{\alpha}}(v)$ has at least four $\alpha$-neighbors in $N^{\alpha}(v)$. Hence there are at least 40 $\alpha$-edges between $N^{\alpha}(v)$ and $N^{\overline{\alpha}}(v)$ in $(T,\lambda)$. 
Since $d^{\alpha}(v)=9$, there exists $u\in N^{\alpha}(v)$ incident to at least $\lceil 40/9\rceil= 5$ such $\alpha$-edges. Moreover, since $d^{\alpha}(v)=9$, Lemma~\ref{lem nbrhood} ensures that $N^{\alpha}(v)$ induces $(SP_9,\square)$, and hence $u$ has four $\alpha$-neighbors in $N^{\alpha}(v)$. Observe also that $v$ is an $\alpha$-neighbor of $u$. Thus, we deduce that $d^\alpha(u)\geq 5+4+1=10$ as desired.
\end{proof}

Let $uv$ be an $\alpha$-edge of $(T,\lambda)$ that is as described in Lemma~\ref{lem uv property}. We now exhibit properties of the neighborhoods of $u$ and $v$ in $(T,\lambda)$.

\begin{lemma}\label{lem a}
Let $A=N^{\alpha}(v)\cap N^{\overline{\alpha}}(u)$. We have $|A|=4$. Moreover, there exists $x\in N^{\overline{\alpha}}(u)$ such that $A=N^{\overline{\alpha}}(u)\cap N^{\overline{\alpha}}(x)$.
\end{lemma}

\begin{proof}
Recall that the signed subgraphs induced by $N^{\alpha}(v)$ and $N^{\overline{\alpha}}(u)$ are both isomorphic to $(SP_9, \square)$, due to Lemma~\ref{lem nbrhood}. Observe that $A$ coincides with the $\overline{\alpha}$-neighborhood of $u$ in $N^{\alpha}(v)$. In particular, since $u\in N^{\alpha}(v)$, $A$ contains exactly four vertices inducing $(K_4, M^{\overline{\alpha}})$ according to Observation~\ref{obs neighborhood of SP9}. 
Furthermore, by Observation~\ref{obs neighborhood of SP9 reverse}, because $A$ induces $(K_4, M^{\overline{\alpha}})$ inside $N^{\overline{\alpha}}(u)$ (which is also sp-isomorphic to $(SP_9, \square)$), there exists $x \in N^{\overline{\alpha}}(u)$ such that $A \subseteq N^{\overline{\alpha}}(x)$. Observe that $A$ is precisely the $\overline{\alpha}$-neighborhood of $x$ in the subgraph induced by $N^{\overline{\alpha}}(u)$. Hence, $A = N^{\overline{\alpha}}(x) \cap N^{\overline{\alpha}}(u)$. 
\end{proof}

We now reach a contradiction by showing that $N^{\alpha}(x)$ has size 9 (and thus induces $(SP_9,\square)$) and contains two disjoint copies of $(K_4,M^\alpha)$, which is impossible. These statements are summarized in the following lemmas.

\begin{lemma}\label{lem bc}
The sets $B = N^{\overline{\alpha}}(u) \setminus (A \cup \{x\})$ and $C=N^{\alpha}(v)\cap N^\alpha(u)$ are disjoint and they both induce $(K_4,M^\alpha)$ in $(T,\lambda)$. Moreover, we have $B=N^{\alpha}(x) \cap N^{\overline{\alpha}}(u)$.
\end{lemma}

\begin{proof}
Since $N^{\overline{\alpha}}(u)$ induces the signed complete graph $SP_9$, $B$ is the set of all neighbors of $x$ in $N^{\overline{\alpha}}(u)$ that are not in $A$, i.e., all the $\alpha$-neighbors of $x$. Hence, $B = N^{\alpha}(x) \cap N^{\overline{\alpha}}(u)$. Observation~\ref{obs neighborhood of SP9} thus yields that $B$ induces $(K_4,M^\alpha)$.

The same argument, applied to the copy of $SP_9$ induced by $N^{\alpha}(v)$, ensures that $C$ also induces $(K_4,M^\alpha)$. Moreover, since $B\subset N^{\overline{\alpha}}(u)$ and $C\subset N^\alpha(u)$, these sets are disjoint.
\end{proof}

\begin{lemma}
$N^{\alpha}(x)$ contains $B\cup C$ and has size 9.
\end{lemma}

\begin{proof}
First observe that, by Lemma~\ref{lem bc}, the vertices of $B$ are $\alpha$-neighbors of $x$. Hence, $B\subset N^\alpha(x)$. 
Now, since $v$ has degree 19, $v$ and $x$ are adjacent and Lemma~\ref{lem common nbrhood} ensures that $N^{\alpha}(v)$ contains at least four $\alpha$-neighbors of $x$. Observe now that $N^\alpha(v)=A\cup C\cup\{u\}$ and that $A\cup\{u\}$ are $\overline{\alpha}$-neighbors of $x$ (by Lemma~\ref{lem a}). Therefore, the four $\alpha$-neighbors of $x$ in $N^{\alpha}(v)$ are precisely the vertices of $C$, i.e. $C= N^{\alpha}(v)\cap N^\alpha(x)\subset N^\alpha(x)$.

We now exhibit 10 $\overline{\alpha}$-neighbors of $x$, which will ensure that $|N^\alpha(x)|=9$. By Lemma~\ref{lem a}, we already know five such neighbors, namely $u$ and the vertices in $A$. Moreover, since $N^\alpha(v)=A\cup C\cup\{u\}$ and $C\subset N^\alpha(x)$, we get that $x\notin N^\alpha(v)$, and, hence, $v$ is another $\overline{\alpha}$-neighbor of $x$. 
Now, by Lemma~\ref{lem common nbrhood}, there are at least four $\overline{\alpha}$-neighbors of $x$ in $N^{\overline{\alpha}}(v)$. Thus, because $x$ has four more  $\overline{\alpha}$-neighbors in $A$ and two more in $\{u, v\}$, $x$ has a total of 10  $\overline{\alpha}$-neighbors. So, we finally deduce that $|N^{\overline{\alpha}}(x)|=10$, and, thus, that $|N^\alpha(x)|=9$.
\end{proof}

\section{Proof of Theorem~\ref{th triangle-free planar bound}}\label{sec 2}

Let $(T, \lambda)$ be a minimal  bound of $\mathcal{P}_4$ of order~$6$. We will show that $(T, \lambda)$ must be isomorphic to $(SP_5^+, \square^+)$ by essentially proving that $(T, \lambda)$ must have very specific properties, converging towards the precise ones that $(SP_5^+, \square^+)$ has.
To do so, we will construct some signed graphs $(H_0,\pi_0)$, $(H_1,\pi_1)$, $\dots$, all being triangle-free and planar, and, thus, 
admitting homomorphisms to $(T, \lambda)$.
These $(H_i,\pi_i)$'s will be constructed gradually, so that each of the $(H_i,\pi_i)$'s allows to deduce more properties of $(T,\lambda)$.

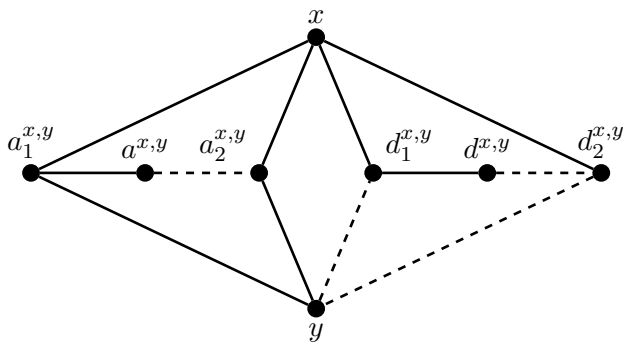
\begin{figure}[!ht]

\centering
\begin{tikzpicture}[inner sep=0.7mm,scale=0.6]

\node[draw, circle, fill, black, line width=1pt](x) at (6.25,3)[label=above:{ $x$}]{};
\node[draw, circle, fill, black, line width=1pt](y) at (6.25,-3)[label=below:{ $y$}]{};

\node[draw, circle, fill, black, line width=1pt](a1) at (0,0)[label=above:{ $a_1^{x,y}$}]{};
\node[draw, circle, fill, black, line width=1pt](a) at (2.5,0)[label=above:{ $a^{x,y}$}]{};
\node[draw, circle, fill, black, line width=1pt](a2) at (5,0)[label=above left:{ $a_2^{x,y}$}]{};
\node[draw, circle, fill, black, line width=1pt](d1) at (7.5,0)[label=above right:{ $d_1^{x,y}$}]{};
\node[draw, circle, fill, black, line width=1pt](d) at (10,0)[label=above:{ $d^{x,y}$}]{};
\node[draw, circle, fill, black, line width=1pt](d2) at (12.5,0)[label=above:{ $d_2^{x,y}$}]{};

\draw[-, line width=1pt,black] (d2) -- (x) -- (a1) -- (a);
\draw[-, line width=1pt,black] (x) -- (a2) -- (y) -- (a1);
\draw[-, line width=1pt,black] (d) -- (d1) -- (x);

\draw[-, line width=1pt,black,dashed] (a) -- (a2);
\draw[-, line width=1pt,black,dashed] (d) -- (d2) -- (y) -- (d1);







\end{tikzpicture}

\caption{The main gadget $(H, \pi)$ used to prove Theorem~\ref{th triangle-free planar bound}. Solid edges are positive edges. Dashed edges are negative edges.
Vertices $x$ and $y$ agree on $a_1^{x,y}$, $a_2^{x,y}$ and disagree on $d_1^{x,y}$, $d_2^{x,y}$.
}\label{fig gadget}
\end{figure}

To construct these $(H_i,\pi_i)$'s, we will mainly use the triangle-free planar signed graph $(H, \pi)$
depicted in Figure~\ref{fig gadget} as a building block.
In what follows, it is important to keep in mind that we deal with the vertices and edges of $(H, \pi)$ using the notation introduced in Figure~\ref{fig gadget}. 

Given a signed graph $(G, \Sigma)$ and one of its vertices $v$, by \textit{pinning
$(H, \pi)$ on $v$} we mean starting from $(G, \Sigma)$, adding a copy of $(H, \pi)$, and identifying the vertex $x$ of 
$(H, \pi)$ with the vertex $v$ of $(G, \Sigma)$. Similarly, for two distinct vertices $u$ and $v$ of $(G, \Sigma)$, by \textit{pinning $(H, \pi)$ on $(u,v)$} we mean starting from $(G, \Sigma)$, adding a copy of 
$(H, \pi)$, identifying the vertex $x$ of 
$(H, \pi)$ with the vertex $u$ of $(G, \Sigma)$,
and similarly identifying $y$ with $v$.
Observe that if $(G, \Sigma)$ is a triangle-free planar signed graph, and $u$ and $v$ are two non-adjacent vertices of $(G, \Sigma)$ belonging to a same face, then the signed graph obtained from $(G, \Sigma)$ by pinning $(H, \pi)$ on $(u,v)$ is also a triangle-free planar signed graph.

Note that the vertices of $(H, \pi)$ are named 
as functions of $x$ and $y$. This will allow us to refer to 
vertices of a copy of $(H, \pi)$ after pinning it to, say, $(u,v)$ of $(G, \Sigma)$, as functions of $u$ and $v$. Since we will deal with larger and larger signed graphs containing multiple copies of $(H, \pi)$, this terminology will allow us to refer to particular vertices in an unambiguous way.

\medskip

We first show that $(T, \lambda)$ must be a signed complete graph. This is done by making use of the following observation. Recall that a negative cycle in a signed graph is a cycle having an odd number of negative edges, while a positive cycle has an even number of negative edges.

\begin{observation}[see e.g.~\cite{nrs2}, Lemma 3.10]\label{obs clique}
Two vertices of a signed graph  have distinct images under every homomorphism if and only if they are adjacent or they are part of a negative $4$-cycle. 
\end{observation}

In the next result, we construct some first triangle-free planar signed graphs, from which we get that $(T, \lambda)$ must indeed be complete.
We start from $(H_0, \pi_0)$ being the signed graph $(H, \pi)$ itself, in which we slightly modify the names of the vertices. That is, we refer to the vertices of $(H_0, \pi_0)$ as in $(H, \pi)$, except that we omit the superscripts (if any). Thus, the vertices $x, y$ retain their name, while the vertices $a_1^{x,y}$, $a^{x,y}$, $a_2^{x,y}$, $d_1^{x,y}$, $d^{x,y}$ and $d_2^{x,y}$ are now, in $(H_0, \pi_0)$, named $a_1$, $a$, $a_2$, $d_1$, $d$ and $d_2$, respectively. 

\begin{lemma}
$(T, \lambda)$ is a signed complete graph. 
\end{lemma}

\begin{proof}
Let $(H_1, \pi_1)$ be the signed graph obtained from $(H_0, \pi_0)$ by pinning $(H, \pi)$ 
on $(x, a)$. 
Note that $(H_1, \pi_1)$ is a triangle-free planar signed graph, and, thus, according to our assumption there exists a homomorphism  $g: (H_1, \pi_1) \rightarrow (T, \lambda)$. 
By Observation~\ref{obs clique}, the vertices $x$, $y$, $a_1$,  $a_2$, $d_1$ and $d_2$ of $(H_1, \pi_1)$ have distinct images in $(T, \lambda)$ under every homomorphism 
$(H_1,\pi_1) \rightarrow (T, \lambda)$. Furthermore, observe that the images of the vertices 
$x$, $a$, $a_1^{x,a}$, $a_2^{x,a}$, $d_1^{x,a}$ and $d_2^{x,a}$ are also distinct. Therefore, since 
$x$, $y$ and $a$ must have distinct images and 
$(T, \lambda)$ has exactly six vertices, 
the images of $a_1^{x,a}$, $a_2^{x,a}$, $d_1^{x,a}$ and $d_2^{x,a}$ must contain the image of $y$. 
In other words, we must have 
$g(y) \in \{g(a_1^{x,a}), g(a_2^{x,a}), g(d_1^{x,a}), g(d_2^{x,a})\}$. 
Therefore, $g(x)$ must be adjacent to $\{g(a_1),g(a_2),g(d_1),g(d_2),g(y)\}$ in $(T,\lambda)$, hence has degree 5.

Next, let $(H_2, \pi_2)$ be the signed graph obtained in the following manner: for each vertex $v$ of $(H_0, \pi_0)$, we glue a copy of $(H_1, \pi_1)$  by identifying the vertex $x$ of $(H_1, \pi_1)$ with the vertex $v$ of $(H_0, \pi_0)$. 
Note that $(H_2, \pi_2)$ is also a triangle-free planar signed graph. Therefore, it admits a homomorphism to 
$(T, \lambda)$. 
By a previous remark, the vertices $x$, $y$, $a_1$,  $a_2$, $d_1$  and $d_2$ must have distinct images by every homomorphism 
$(H_2,\pi_2) \rightarrow (T, \lambda)$. Hence, there must be six distinct vertices of degree $5$ in $(T, \lambda)$, which thus must be complete.
\end{proof}

Let now $(H_3, \pi_3)$ be the signed graph obtained from $(H_0, \pi_0)$ by pinning four $(H, \pi)$'s  on $(x, a)$, $(x, d)$, $(y, a)$ and $(y, d)$, respectively.
Note that $(H_3, \pi_3)$ is a triangle-free planar signed graph, and thus it admits a homomorphism to $(T, \lambda)$. 
In what follows, we need to understand better the different types of homomorphisms of $(H_3, \pi_3)$ to $(T, \lambda)$.

Let $f$ be a homomorphism of $(H_3, \pi_3)$ to $(T, \lambda)$.
For convenience, suppose that $V(T) = \{1,2,3,4,5,6\}$. 
By Observation~\ref{obs clique}, we know that, in $(H_3, \pi_3)$, the vertices $x$, $y$, $a_1$, $a_2$, $d_1$ and $d_2$ have distinct images by $f$.  Without loss of generality, we may assume that these images by $f$ are as displayed in the following table:

\medskip

\begin{center}
\begin{tabular}{|c|c|c|c|c|c|}
\hline
  $f(x)$  & $f(y)$ & $f(a_1)$ & $f(a_2)$ & $f(d_1)$ & $f(d_2)$ \\
\hline
 $1$  & $2$ & $3$ & $4$ & $5$ & $6$ \\
\hline
\end{tabular}   
\end{center}

\medskip

Furthermore,  by Observation~\ref{obs clique} we know that 
$f(a) \in \{5,6\}$ and $f(d) \in \{3,4\}$. Thus, without loss of generality, we may also assume  $f(a)=5$, which implies $f(d)=3$:

\medskip

\begin{center}
\begin{tabular}{|c|c|}
\hline
  $f(a)$  & $f(d)$  \\
\hline
 $5$  & $3$  \\
\hline
\end{tabular}   
\end{center}
Note that this may require to switch some vertices among $a$ and $d$, but in that case we can relabel some vertices of the four pinned copies of $(H,\pi)$ in $(H_3,\pi_3)$ and keep the original signature of $(H_3,\pi_3)$. 

\medskip

Now, let us focus on the copy of $(H, \pi)$ in $(H_3, \pi_3)$ that was pinned on $(x, a)$.
Notice, by Observation~\ref{obs clique}, that $f(x), f(a)$, $f(a_1^{x,a})$, $f(a_2^{x,a})$, $f(d_1^{x,a})$ and $f(d_2^{x,a})$ are pairwise distinct, that
$f(a_1^{x,a}), f(a_2^{x,a}) \in \{2,3,6\}$ (since they agree on vertices $1$ and~$5$),  and that
$f(d_1^{x,a}), f(d_2^{x,a}) \in \{2,4,6\}$ (since they disagree on vertices $1$ and~$5$).
Therefore, either $f(a_1^{x,a})=3$ or $f(a_2^{x,a}) = 3$ and, similarly,
either $f(d_1^{x,a})=4$ or $f(d_2^{x,a}) = 4$.  
As we have not assumed that
$(H_3, \pi_3)$ is embedded in the plane in a specific way, due to the symmetric structure of the graph, we may assume without loss of generality 
 $f(a_1^{x,a})=3$ and $f(d_2^{x,a}) = 4$.
Reasoning similarly on the other copies of $(H, \pi)$, we may suppose that we have the following images by $f$:

\medskip

\begin{center}
\begin{tabular}{|c|c|c|c|c|c|c|c|}
\hline
$f(a_1^{x,a})$  & $f(d_2^{x,a})$ & $f(a_1^{x,d})$ & $f(d_2^{x,d})$ \\
\hline
 $3$  & $4$ & $5$ & $6$ \\
\hline
\hline
 $f(a_1^{y,a})$ & $f(d_2^{y,a})$ & $f(a_1^{y,d})$ & $f(d_2^{y,d})$ \\
\hline
 $3$ & $4$ & $6$ & $5$ \\
\hline
\end{tabular}   
\end{center}

\medskip

We now analyze the possible images by $f$ for some of the remaining vertices of $(H_3, \pi_3)$. First, note that, by Observation~\ref{obs clique}, we have the following:


\begin{observation}\label{observation:assumpt}
By Observation~\ref{obs clique}, we have:

\begin{itemize}
    \item $\{f(a_2^{y,a}), f(d_1^{y,a})\}= \{1,6\}$,
    \item $\{f(a_2^{x,d}), f(d_1^{x,d})\}=\{2,4\}$,
    \item $\{f(a_2^{y,d}), f(d_1^{y,d})\}=\{1,4\}$,
    \item $\{f(a_2^{x,a}), f(d_1^{x,a})\}= \{2,6\}$.
\end{itemize}


\end{observation}

Regarding the first item in Observation~\ref{observation:assumpt}, there are two possibilities for $f$, namely either $(f(a_2^{y,a}), f(d_1^{y,a}))= (1,6)$, or conversely $(f(a_2^{y,a}), f(d_1^{y,a}))= (6,1)$.
In the next two lemmas, we analyze the consequences on $f$ of being in one case or the other.

\begin{lemma}\label{lem f1}
If $f(a_2^{y,a})=1$ and $f(d_1^{y,a}) = 6$, then we have the following images by $f$:

\medskip

\begin{center}
\begin{tabular}{|c|c|c|c|c|c|c|c|}
\hline
$f(a_2^{x,a})$  & $f(d_1^{x,a})$ & $f(a_2^{x,d})$ & $f(d_1^{x,d})$ \\
\hline
 $6$  & $2$ & $2$ & $4$ \\
\hline
\hline
 $f(a_2^{y,a})$ & $f(d_1^{y,a})$ & $f(a_2^{y,d})$ & $f(d_1^{y,d})$ \\
\hline
 $1$ & $6$ & $1$ & $4$ \\
\hline
\end{tabular}   
\end{center}
\end{lemma}

\begin{proof}
If  $f(d_1^{x,d})=2$, then the positive cycle $a_2^{y,a} y a_1^{y,a} a a_2^{y,a}$ and the negative cycle $x d_1^{x,d} d a_1^{x,d} x$ of $(H_3, \pi_3)$ have the same image $12351$ by $f$, which is a contradiction. 
Therefore, $f(a_2^{x,d})=2$ and $f(d_1^{x,d})=4$. Also, if  $f(a_2^{y,d})=4$, then the negative cycle
$ x d_1^{x,d} d a_2^{y,d} y a_2 x$ has image $1434241$ by $f$, which is a positive closed walk in $(T, \lambda)$, a contradiction.  From this, we deduce that $f(a_2^{y,d})=1$ and $f(d_1^{y,d})=4$. 
Finally, if $f(a_2^{x,a})=2$, then the positive cycle $x a_2^{x,a} a a_1^{x,a} x$ and the negative cycle $a_2^{y,d} y d_2^{y,d} d a_2^{y,d}$ have the same image $12531$ by $f$, a contradiction. Therefore, $f(a_2^{x,a})=6$ and $f(d_1^{x,a})=2$. 
\end{proof}

\begin{lemma}\label{lem f2}
If $f(a_2^{y,a})=6$ and $f(d_1^{y,a}) = 1$, then we have the following images by $f$:

\medskip

\begin{center}
\begin{tabular}{|c|c|c|c|c|c|c|c|}
\hline
$f(a_2^{x,a})$  & $f(d_1^{x,a})$ & $f(a_2^{x,d})$ & $f(d_1^{x,d})$ \\
\hline
 $2$  & $6$ & $4$ & $2$ \\
\hline
\hline
 $f(a_2^{y,a})$ & $f(d_1^{y,a})$ & $f(a_2^{y,d})$ & $f(d_1^{y,d})$ \\
\hline
 $6$ & $1$ & $4$ & $1$ \\
\hline
\end{tabular}   
\end{center}
\end{lemma}

\begin{proof}
If  $f(a_2^{x,d})=2$, then the positive cycle $x a_2^{x,d} d a_1^{x,d} x$ and the negative cycle $d_1^{y,a} y a_1^{y,a} a d_1^{y,a}$ of $(H_3, \pi_3)$ have the same image $12351$ by $f$, which is not possible. 
Therefore, $f(a_2^{x,d})=4$ and $f(d_1^{x,d})=2$. 
Now, if  $f(d_1^{y,d})=4$, then the negative cycle 
$x a_2^{x,d} d d_{1}^{y,d} y a_2 x$ has image
$1434241$ by $f$, which is a positive closed walk in $(T, \lambda)$, a contradiction from which we deduce $f(a_2^{y,d})=4$ and $f(d_1^{y,d})=1$. 
Similarly, if $f(a_2^{x,a})=6$, then the negative cycle
$ x d_2 y a_2^{y,a} a a_2^{x,a} x$ has image 
$1626561$ by $f$, which is a positive closed walk in $(T, \lambda)$, a contradiction.
Then we deduce that $f(a_2^{x,a})=2$ and $f(d_1^{x,a})=6$. 
\end{proof}

From Lemmas~\ref{lem f1} and~\ref{lem f2}, we get that there are, thus far, two possible partial extensions for $f$. We denote by $f_1$ the one described in the statement of Lemma~\ref{lem f1}, and by $f_2$ the one described in the statement of Lemma~\ref{lem f2}.

\medskip


Let $\{i_1, \dots,  i_6\} = \{1, \dots, 6\}$. 
In the signed graph $(T, \lambda)$, if two vertices $i_1,i_2$ agree on
two vertices $i_3, i_4$ and disagree on $i_5, i_6$, then we say that 
$\{i_1, i_2\}$ is a \textit{splitter} that yields two \textit{teams} $\{i_3, i_4\}$ and $\{i_5, i_6\}$. Naturally, in this case, $i_3$ is in the same team as 
$i_4$, that is opposite to the team of $i_5$ and $i_6$. 
Observe that no matter how we switch vertices in $(T, \lambda)$, the pair $\{i_1, i_2\}$ 
remains a splitter yielding the same two teams. 
Upon switching vertices, it may happen that 
$i_1, i_2$ get to disagree on $i_3, i_4$ and to agree on $i_5, i_6$ -- but the fact that $\{i_1, i_2\}$ yields teams $\{i_3, i_4\}$ and $\{i_5, i_6\}$ cannot be lost. 

Having a closer look, in $(H_3, \pi_3)$, at the images by $f$, observe that $x$ and $y$ must agree on $a_1$ and $a_2$ and disagree on $d_1$ and $d_2$. The images by $f$ of $x$ and $y$  thus imply that, in $(T, \lambda)$, the pair $\{1,2\}$ is a splitter yielding teams $\{3,4\}$ and $\{5,6\}$. 
Moreover, because $f(a) = 5$ and there is a copy of $(H, \pi)$ pinned to $(x,a)$, 
then, in $(T, \lambda)$, the pair $\{1,5\}$ must be a splitter. Similarly, because $f(d) = 3$, the pair $\{1,3\}$ must also be a splitter. Thus, vertex $1$ is part of at least three distinct splitters. 
We actually need to show something stronger.

\begin{lemma}
Every vertex of $(T,\lambda)$ is part of at least four distinct splitters. 
\end{lemma}

\begin{proof}
Let $(H_4, \pi_4)$ be the triangle-free planar signed graph obtained by pinning one copy of $(H, \pi)$ to each of the eight pairs $(a_1^{x,a}, a_2^{x,a})$, 
$(d_1^{x,a}, d_2^{x,a})$, 
$(a_1^{x,d}, a_2^{x,d})$,
$(d_1^{x,d}, d_2^{x,d})$,
$(a_1^{y,a}, a_2^{y,a})$, 
$(d_1^{y,a}, d_2^{y,a})$, 
$(a_1^{y,d}, a_2^{y,d})$ and
$(d_1^{y,d}, d_2^{y,d})$ of vertices of $(H_3, \pi_3)$.
Consider an extension of $f$ to $(H_4, \pi_4)$.
Note that if $f$ is extended so that it matches $f_1$, then the copy of $(H, \pi)$ pinned on 
$(a_1^{y,d}, a_2^{y,d})$ implies that $\{1,6\}$ is a splitter. If 
$f$ is extended so that it matches $f_2$, then the copy of $(H, \pi)$ pinned on 
$(d_1^{y,a}, d_2^{y,a})$ implies that $\{1,4\}$ is a splitter.
Earlier, we have already pointed out that  $\{1,2\}, \{1,3\}$ and $\{1,5\}$ are splitters. 
Therefore, because $(H_4, \pi_4)$ verifies $f(x) = 1$, we get that vertex $1$ must be part of at least four splitters.

Let now $(H_5, \pi_5)$ be the triangle-free planar signed graph obtained by starting from $(H_0, \pi_0)$ and, for each of its vertices $u$, adding a copy of $(H_4, \pi_4)$ and identifying $u$ and the vertex $x$ of that copy.
Then, for every 
homomorphism $(H_5, \pi_5) \rightarrow (T, \lambda)$ and for every $i \in V(T)$, there is a copy of $(H_4, \pi_4)$ in $(H_5, \pi_5)$ for which the image of $x$ is $i$.  This completes the proof. 
\end{proof}

In what follows, we prove that if some vertex of $(T, \lambda)$ is part of five distinct splitters, 
then $(T, \lambda)$ must be isomorphic to $(SP_5^+, \square^+)$, in which case we are done.

\begin{lemma}
If a vertex of $(T, \lambda)$ is part of five distinct splitters, 
then $(T, \lambda)$ is isomorphic to $(SP_5^+, \square^+)$. 
\end{lemma}

\begin{proof}
Without loss of generality, assume that, in $(T, \lambda)$, vertex $6$ is part of 
five distinct splitters. Switch the $-$-neighbors of vertex $6$ so that all its incident edges get positive.
Because the set $\{i,6\}$ is a splitter for every $i \in \{1,2,3,4,5\}$, we deduce that every vertex $i$ must be incident to exactly two positive edges and two negative edges in 
$(T - 6, \lambda)$, the signed graph obtained from
$(T, \lambda)$ by deleting vertex $6$. Then, the vertices in $\{1,2,3,4,5\}$ and their incident positive edges must induce a $2$-regular graph. Since the only $2$-regular (simple) graph of order~$5$ is the $5$-cycle, we get the desired conclusion.
\end{proof}

Now assume that no vertex of $(T, \lambda)$ is part of five distinct splitters. We prove that, under that assumption, $(T, \lambda)$ must be one of two possible signed graphs, $(K_6, M)$ and $(K_6, \overline{M})$, defined as follows. Let $M$ be a perfect matching of $K_6$, the complete graph of order~$6$.
The signed graph $(K_6, M)$ is the signed $K_6$ in which the set of negative edges is precisely $M$. The signed graph $(K_6, \overline{M})$ is the signed $K_6$ in which the set of negative edges is the set $\overline{M}=E(K_6) \setminus M$ of the edges that are not in $M$.

\begin{lemma}
If no vertex of $(T, \lambda)$ is part of five distinct splitters, 
then $(T, \lambda)$ is isomorphic to $(K_6, M)$ or $(K_6, \overline{M})$. 
\end{lemma}

\begin{proof}
In this case, each vertex of $(T, \lambda)$ is part of exactly four distinct splitters. 
Let us switch the $-$-neighbors of vertex $6$ to make all its incident edges positive. Let $(T - 6, \lambda)$ be the signed graph 
obtained from $(T, \lambda)$ by deleting vertex $6$.
Note that the subgraph $T^+$ induced by the positive edges of $(T - 6, \lambda)$ must have exactly four vertices of degree~$2$. By the Handshaking Lemma, the fifth vertex $j$ must then have even degree, hence has degree $0$ or $4$. 
If $j$ has degree $0$, then $T^+$ is the disjoint union of a singleton vertex and a $4$-cycle. In this case, by switching $j$ in $(T, \lambda)$ we get the signed graph $(K_6, M)$. 
Now, if $j$ has degree $4$, then $T^+$ is the $1$-clique-sum of two $3$-cycles. In this case, by switching vertex $6$ and $j$ in $(T, \lambda)$, we obtain the signed graph 
$(K_6, \overline{M})$. 
\end{proof}

We complete the proof by showing that it is actually not possible for $(T, \lambda)$ to be isomorphic to $(K_6, M)$ or to $(K_6, \overline{M})$, a contradiction with the previous lemma.

\begin{lemma}
$(T, \lambda)$ cannot be isomorphic to $(K_6, M)$ or $(K_6, \overline{M})$.
\end{lemma}

\begin{proof}
Note that if $(T, \lambda)$ is isomorphic to $(K_6, M)$ or $(K_6, \overline{M})$, then, for every vertex $i$ of $(T, \lambda)$, there exists exactly one other vertex $j$
such that $\{i, j\}$ is not a splitter. 
Let us consider the two possibilities, $f_1$ and $f_2$, for $f$ to be extended in $(H_3, \pi_3)$.

First, assume that $f$ is partially extended as $f_1$. 
The three sets (of cardinality $2$) of vertices of $(T, \lambda)$ that are not splitters are $\{1,4\}$, $\{2, 6\}$ and $\{3,5\}$. Therefore, if $(T, \lambda)$ is isomorphic to $(K_6, M)$, then its three negative edges are $14$, $26$ and $35$. Analogously, if $(T, \lambda)$ is isomorphic to $(K_6, \overline{M})$, then its three positive edges are $14$, $26$ and $35$.
Now, looking at the structure of $(K_6, M)$ or $(K_6, \overline{M})$, the splitter
$\{1,3\}$ yields the two teams $\{2,6\}$ and $\{4,5\}$. 
Let us now look further at the images of the vertices of $(H_3, \pi_3)$ by $f_1$. We know that $f_1(x) = 1$ and $f_1(d) = 3$. Moreover, we know that 
$x$ and $d$ agree on $a_1^{x,d}$ and $a_2^{x,d}$ and disagree on $d_1^{x,d}$ and $d_2^{x,d}$. 
Because $f_1(a_1^{x,d}) = 5$, $f_1(a_2^{x,d}) = 2$, 
$f_1(d_1^{x,d}) = 4$ and $f_1(d_2^{x,d}) = 6$, we can conclude that the splitter $\{1,3\}$ yields the two teams $\{2,5\}$ and $\{4,6\}$, which is a contradiction. Thus if $f$ is extended as $f_1$, then $(T, \lambda)$ must be isomorphic to $(SP_5^+, \square^+)$. 

Second, assume that $f$ is partially extended as $f_2$.
In this case, the three sets (of cardinality~$2$) of vertices of $(T, \lambda)$ that are not splitters are $\{1,6\}$, $\{2, 4\}$ and $\{3,5\}$.
This implies that the splitter $\{1,3\}$ yields the two teams $\{2,4\}$ and $\{5,6\}$.  However, in $(H_3, \pi_3)$,  we have
$f_2(x)=1$, $f_2(d) =3$, 
$f_2(a_1^{x,d}) = 5$, $f_2(a_2^{x,d}) = 4$, 
$f_2(d_1^{x,d}) = 2$ and $f_2(d_2^{x,d}) = 6$.
From these images, we conclude that 
the splitter $\{1,3\}$ yields the two teams $\{4,5\}$ and $\{2,6\}$, which is a contradiction.  Thus if $f$ is extended as $f_2$, then, again, $(T, \lambda)$ must be isomorphic to $(SP_5^+, \square^+)$. 
\end{proof}

\section{Proof of Theorem~\ref{th Kn-minor-free}}\label{sec 3}

We start by proving the upper bounds, which we do by exploiting existing connections between signed graphs and acyclic colorings.
Recall that an \textit{acyclic coloring} of an undirected graph $G$ is a proper vertex-coloring such that the subgraph induced by any two distinct colors is acyclic, i.e., is a forest.
The \textit{acyclic chromatic number} $\chi_a(G)$ of $G$ is the minimum $k$ such 
that $G$ admits an acyclic $k$-coloring. 

\begin{proof}[Proof of Theorem~\ref{th Kn-minor-free}(i).]
It was proved in~\cite{KnminorVSacyclic} 
that $\chi_a(G) \leq 5 {n-1 \choose 2}$ holds for every graph $G \in \mathcal{F}_n$. 
Furthermore, given a signed graph $(G, \sigma)$,
it is also known that if $\chi_a(G) \leq k$, then
$\chi_s((G, \sigma)) \leq k2^{k-2}$ (see~\cite{DBLP:journals/jgt/OchemPS17})
and $\CHISP((G, \sigma)) \leq k2^{k-1}$ (see~\cite{alon1998homomorphisms}). 
Combining these bounds yields the desired upper bounds.
\end{proof}

We say that a family $\mathcal{F}$ of graphs is \textit{complete} if for every 
finite collection $\mathcal{C} = \{G_1, \dots, G_t\}$ of graphs from $\mathcal{F}$, 
the graph obtained by taking the disjoint union of all graphs of $\mathcal{C}$ also belongs to $\mathcal{F}$.

\begin{lemma}\label{lemma:complete-family}
Every complete family $\mathcal{F}$ of graphs has an 
sp-bound of order~$\CHISP(\mathcal{F})$  and a bound of order~$\chi_s(\mathcal{F})$.  
\end{lemma}

\begin{proof}
Suppose $\mathcal{F}$ does not have an sp-bound of order~$n=\CHISP(\mathcal{F})$. 
Let $\mathcal{S}$ be the set of all signatures of $K_n$.  Since $\mathcal{F}$ does not have any sp-bound 
of order~$n$, for each $\pi \in \mathcal{S}$ there exists a $(G_{\pi}, \sigma_{\pi})$ 
that does not admit an sp-homomorphism to $(K_n, \pi)$. 
Let $(G, \sigma)$ be the signed graph containing $(G_{\pi}, \sigma_{\pi})$ as a subgraph for all $\pi \in \mathcal{S}$. That is, $(G, \sigma)$ is the disjoint union of all possible $(G_\pi,\sigma_\pi)$'s, where $\pi$ runs across $\mathcal{S}$.
Observe that
$G \in \mathcal{F}$. 
Furthermore, note that $(G, \sigma)$ does not admit an sp-homomorphism to $(K_n, \pi)$ for any $\pi \in \mathcal{S}$. Thus $\CHISP((G, \pi)) > n$, a contradiction. 

The proof for the existence of a bound of order $\chi_s(\mathcal{F})$ is similar. 
\end{proof}

With Lemma~\ref{lemma:complete-family} on hand, we can now prove the second part of Theorem~\ref{th Kn-minor-free}.

\begin{proof}[Proof of Theorem~\ref{th Kn-minor-free}(ii).]
We know from~\cite{Montejano_2ec, DBLP:journals/jgt/OchemPS17} that the result holds for $n=1, 2, 3, 4, 5$. 
We prove the result for larger values of $n$ by induction.
Suppose that the result holds for all $n \leq t$ where $t\ge 5$ is odd. We show that the result holds for $n = t+1$ and $n=t+2$.

We first prove the result for $n=t+1$.
Consider the following construction. Given a signed graph $(G, \sigma)$, take two disjoint copies $(G_1, \sigma_1)$ and $(G_2, \sigma_2)$ of $(G, \sigma)$, add a new vertex $\infty$, and 
make every vertex of $(G_1, \sigma_1)$ adjacent to $\infty$ via a positive edge and every vertex of 
$(G_2, \sigma_2)$ adjacent to $\infty$ via a negative edge. 
We denote the so-obtained signed graph by $(G^*, \sigma^*)$.
Observe that 
\begin{equation}\label{eqn 2n+1}
\CHISP((G^*, \sigma^*)) = 2 \CHISP((G, \sigma)) +1.
\end{equation}
Indeed, if $(G,\sigma)\xrightarrow{sp} (H,\pi)$ with $|V(H)|=\CHISP((G,\sigma)) $, then $(G^*,\sigma^*)\xrightarrow{sp} (H^*,\pi^*)$, which ensures that $\CHISP((G^*,\sigma^*)) \leq |V(H^*)|=2\CHISP((G,\sigma))+1$. For the reverse inequality, assume that there is an sp-homomorphism $f:(G^*,\sigma^*)\xrightarrow{sp} (H,\pi)$. Then one of the copies of $(G,\sigma)$ in $(G^*,\sigma^*)$ is mapped by $f$ in $H[N^+(f(\infty))]$, and the other in $H[N^-(f(\infty))]$. The inequality then follows from the fact that at least one of these subgraphs has order at most $\frac{|V(H)\setminus\{\infty\}|}{2}=\frac{\CHISP((G^*,\sigma^*))-1}{2}$.

Let $(H_t, \pi_t)$ be a signed graph with $\CHISP((H_t, \pi_t)) \geq \frac{2^{t+1}-4}{3}$, where 
$H_t \in \mathcal{F}_t$. Let us set $(H_{t+1}, \pi_{t+1}) = (H_t^*, \pi_t^*)$. Note that 
$H_{t+1} \in \mathcal{F}_{t+1}$;
therefore, by Equation~(\ref{eqn 2n+1}), we have  
\begin{align*}
\CHISP((H_{t+1}, \pi_{t+1})) &= 2 \CHISP((H_t, \pi_t)) +1 \\
&\geq 2\cdot\frac{2^{t+1}-4}{3} +1\\
                   &= \frac{2^{t+2}-8+3}{3} = \frac{2^{(t+1)+1}-5}{3}.
\end{align*}
This example implies the lower bound for the case $n = t+1$. 

\medskip

We now prove the result for $n=t+2$. 
First of all, consider the signed graph $(H_{t+2}, \pi_{t+2}) = (H_{t+1}^*, \pi_{t+1}^*)$. 
By Equation~(\ref{eqn 2n+1}) we have  
\begin{align*}
\CHISP((H_{t+2}, \pi_{t+2})) &= 2 \CHISP((H_{t+1}, \pi_{t+1})) +1 \\
&\geq 2\cdot\frac{2^{(t+1)+1}-5}{3} +1\\
                   &= \frac{2^{t+3}-10+3}{3} = \frac{2^{(t+2)+1}-4}{3}-1.
\end{align*}
Since $H_{t+2} \in \mathcal{F}_{t+2}$, the result will hold if we can prove that we cannot have equality in the second line of the equation above. Thus, assume the contrary, i.e., 
$$\CHISP((H_{t+1}, \pi_{t+1})) = \frac{2^{(t+1)+1}-5}{3} \text{ ~~~and~~~ } \CHISP((H_t, \pi_t)) = \frac{2^{t+1}-4}{3}.$$

Now consider the following construction, similar to the ones depicted on Figures~\ref{fig copiesv} and~\ref{fig copiese}. Take $(H_{t+2}, \pi_{t+2})$ and $|V(H_{t+2})|$ copies of $(H_{t+1}^*, \pi_{t+1}^*)$. After that, for every vertex $v \in V(H_{t+2})$,  take a copy of $(H_{t+1}^*, \pi_{t+1}^*)$ and
 identify $v$ with the vertex $\infty$. We call the resulting graph $(H'_{t+2}, \pi'_{t+2})$.
We further enhance this construction as follows. Take the disjoint union of $(H'_{t+2}, \pi'_{t+2})$ and of
$4|E(H'_{t+2})|$ copies of $(H_t, \pi_t)$. Then, for every edge $e = uv \in E(H'_{t+2})$ 
and every pair $(\alpha, \beta) \in \{+,-\}^2$, take a copy of $(H_t, \pi_t)$,
make its vertices adjacent to $u$ through $\alpha$-edges and to $v$ through $\beta$-edges. We denote the resulting graph by $(H''_{t+2}, \pi''_{t+2})$.
 
If $\CHISP(\mathcal{F}_{t+2}) < \frac{2^{(t+2)+1}-4}{3}$ (contradicting the statement of the result we want to prove), then we must have 
$$\frac{2^{(t+2)+1}-4}{3}-1>\CHISP(\mathcal{F}_{t+2}) \geqslant \CHISP((H''_{t+2}, \pi''_{t+2})) \geqslant \CHISP((H_{t+2},\pi_{t+2}))= \frac{2^{(t+2)+1}-4}{3} -1.$$ 
This implies that there exists a signed graph $(T, \lambda)$ of order $\frac{2^{(t+2)+1}-4}{3} -1$ such that $(H''_{t+2}, \pi''_{t+2}) \xrightarrow{sp} (T, \lambda)$. Let 
$f: (H''_{t+2}, \pi''_{t+2}) \xrightarrow{sp} (T, \lambda)$ be an sp-homomorphism. 
 Note that $f$ is surjective, since 
$\CHISP((H''_{t+2}, \pi''_{t+2})) = \frac{2^{(t+2)+1}-4}{3} -1$. 
As we also have  $\CHISP((H_{t+2}, \pi_{t+2})) = \frac{2^{(t+2)+1}-4}{3} -1$,
we get that the vertices of the original $(H_{t+2}, \pi_{t+2})$ contained in 
$(H''_{t+2}, \pi''_{t+2})$ as a subgraph also map onto the vertices of $(T, \lambda)$. 
From this, we may infer that every vertex $x$ of $(T, \lambda)$ has a copy of $(H_{t+1}, \pi_{t+1})$
mapped to its $\alpha$-neighborhood by $f$ for every $\alpha \in \{+,-\}$. Thus every vertex $x$ of 
$(T, \lambda)$ must have at least $\CHISP((H_{t+1}, \pi_{t+1})) = \frac{2^{(t+1)+1}-5}{3}$ $\alpha$-neighbors for every $\alpha \in \{+,-\}$. Since $T$ has exactly 
$$\frac{2^{(t+2)+1}-4}{3} -1 = 2 \cdot \frac{2^{(t+1)+1}-5}{3} +1$$ vertices, every vertex of $(T, \lambda)$ must thus have exactly 
$\frac{2^{(t+1)+1}-5}{3}$ $\alpha$-neighbors for every $\alpha \in \{+,-\}$. Hence,  $(T, \lambda)$ is a  complete signed graph.
Furthermore,  for every edge $e = uv \in E(H_{t+2})$ of the original copy contained in 
$(H''_{t+2}, \pi''_{t+2})$ as a subgraph  
and for every pair $(\alpha, \beta) \in \{+,-\}^2$, there is a copy of $(H_t, \pi_t)$ contained in the subgraph induced by $N^{\alpha}(u) \cap N^{\beta}(v)$. Thus, in particular, for any distinct pair of vertices  $x, y$ of $(T, \lambda)$, $N^\alpha(x) \cap N^\beta(y)$ contains at least $\CHISP((H_t, \pi_t)) = \frac{2^{t+1}-4}{3}$ vertices for every $(\alpha, \beta) \in \{+, -\}^2$. 
 \medskip

To reach a contradiction, we count in two ways the number of $+$-edges between $A$ and $B$. Let $x$ be a vertex of $(T, \lambda)$. We already know that the $+$-neighborhood $A$ of 
$x$ in $(T, \lambda)$ contains exactly  $\frac{2^{(t+1)+1}-5}{3}$ vertices and the 
$-$-neighborhood $B$ of 
$x$ in $(T, \lambda)$ contains exactly  $\frac{2^{(t+1)+1}-5}{3}$ vertices. 
Moreover, every vertex $y$ in $A$  has exactly $\frac{2^{t+1}-4}{3}$ $\alpha$-neighbors 
 in $A$ for every $\alpha \in \{+,-\}$. 
Note that $y$ already has one $+$-neighbor, $x$, and $\frac{2^{t+1}-4}{3}$ $+$-neighbors in $A$. 
Hence, $y$ must have exactly 
$$\frac{2^{(t+1)+1}-5}{3} - \frac{2^{t+1}-4}{3} -1 = \frac{2^{t+1}-4}{3}$$
 $+$-neighbors in $B$. 
Thus, there are exactly $\frac{(2^{(t+1)+1}-5)(2^{t+1}-4)}{9}$ $+$-edges between the sets 
$A$ and $B$. 
Similarly, every vertex $z$ in $B$  has exactly $\frac{2^{t+1}-4}{3}$ $\alpha$-neighbors 
in $A$ for every $\alpha \in \{+,-\}$. 
Note that $z$ already has $\frac{2^{t+1}-4}{3}$ $+$-neighbors in $B$. 
Hence, it must have exactly 
$$\frac{2^{(t+1)+1}-5}{3} - \frac{2^{t+1}-4}{3} = \frac{2^{t+1}-1}{3}$$
$+$-neighbors in $B$. 
Thus, there are exactly $\frac{(2^{(t+1)+1}-5)(2^{t+1}-1)}{9}$ 
$+$-edges between the sets $A$ and $B$. This is a contradiction with the previous counting, which implies that $(T, \lambda)$ cannot exist. This concludes the proof.
\end{proof}

This leaves us with proving the very last part of Theorem~\ref{th Kn-minor-free}.

\begin{proof}[Proof of Theorem~\ref{th Kn-minor-free}(iii).]
If $n$ is even, Theorems~\ref{prop 2-ec signed} and~\ref{th Kn-minor-free}(ii) give that
$$2\chi_s((G,\sigma))\geq \CHISP((G,\sigma)) \geq \frac{2^{n+1}-5}{3}.$$
Hence, because $\chi_s((G,\sigma))$ is an integer, we get
$$\chi_s((G,\sigma))\geq \left\lceil \frac{2^{n+1}-5}{6}\right\rceil  = \left\lceil \frac{2^n-1}{3}-\frac{1}{2}\right\rceil = \frac{2^n-1}{3}$$ 
since $n$ is even. The case when $n$ is odd is similar.
\end{proof}


\section{Proof of Theorem~\ref{th signed cubic+}}\label{sec 4}

Throughout this section, we say that each of the two signed graphs $(K_4,\emptyset)$ (having positive edges only) and $(K_4,E(K_4))$ (having negative edges only) is a \textit{bad $K_4$}, while every other signature of $K_4$ gives a \textit{good $K_4$}. 

We first observe that $(SP_5^+,\square^+)$ contains a copy of each good $K_4$.
\begin{observation}\label{obs some K4}
If $(K_4,\Sigma)$ is not bad, then $(K_4, \Sigma) \rightarrow (SP_5^+, \square^+)$.
\end{observation}

\begin{proof}
Given a signature $\Sigma$ of $K_4$, one can switch some vertices to obtain an equivalent signature $\Sigma'$ in which some vertex $v$ has its three incident edges being positive. If $(K_4,\Sigma)$ is not bad, then the signed graph obtained by deleting $v$ from $(K_4, \Sigma')$ does not have only positive edges or only negative edges and hence can be found in $(SP_5,\square)$. Therefore $(SP_5^+, \square^+)$ contains $(K_4, \Sigma')$ as a subgraph, where $v$ is mapped to $\infty$. 
\end{proof}

In this section, we want to show that the family of all signed subcubic graphs with no bad $K_4$ as a connected component, admits a homomorphism to $(SP_5^+, \square^+)$. The proof is by contradiction.
Suppose there exists a signed subcubic graph with no bad $K_4$ as a connected component, that does not admit a homomorphism to $(SP_5^+, \square^+)$. We focus on $(G, \sigma)$, a counterexample that is minimal in terms of order. 
That is, every signed subcubic graph with fewer vertices than $(G, \sigma)$ admits a homomorphism to $(SP_5^+, \square^+)$.
Our goal is to show that $(G, \sigma)$ cannot exist, a contradiction.
This is done by investigating properties of $(G, \sigma)$,
and considering homomorphisms to $(SP_5^+, \square^+)$ (depicted in Figure~\ref{fig Paley}(a)). 

\medskip

By minimality, we observe that $(G, \sigma)$ is connected.
Also, $G \neq K_4$ (by Observation~\ref{obs some K4}).
We start off by showing that $(G, \sigma)$ cannot have cut-vertices.

\begin{lemma}\label{lemma:connected}
$(G, \sigma)$ is $2$-connected. 
\end{lemma}

\begin{proof}
Assume that $(G, \sigma)$ has a cut-vertex $v$.
Then, removing $v$ from $(G, \sigma)$ results in at least two connected components. Assume that
$(G_1, \sigma_1)$  is one such connected component, and $(G_2, \sigma_2)$ is the disjoint union of all the other connected components. 
Let $(G_1', \sigma_1')$ be the signed graph obtained by putting the vertex $v$ back in $(G_1, \sigma_1)$, and let $(G_2', \sigma_2')$ be the signed graph obtained by putting the vertex $v$ back in $(G_2, \sigma_2)$. 
Note that none of these two signed graphs is cubic, and, thus, none of them can be a bad $K_4$.
By minimality of $(G, \sigma)$, there are 
$f_1: (G_1', \sigma_1') \rightarrow (SP_5^+, \square^+)$
and $f_2: (G_2', \sigma_2') \rightarrow (SP_5^+, \square^+)$. 
Due to the vertex-transitivity of $(SP_5^+, \square^+)$, we may assume  $f_1(v) = f_2(v)$. Now, combining $f_1$ and $f_2$ yields a homomorphism of
$(G, \sigma)$ to $(SP_5^+, \square^+)$, a contradiction. 
\end{proof}

Through the next result, we aim at reducing $(G, \sigma)$ to a cubic graph. Note that $(G,\sigma)$ has no vertex of degree 1 since it is 2-connected. 


\begin{lemma}\label{lem deg 3}
$(G, \sigma)$ does not contain a vertex of degree $2$. 
\end{lemma}

\begin{proof}
Suppose the contrary, i.e., assume that $(G, \sigma)$  contains a degree-$2$ vertex $u$ with neighbors $v$ and $w$.
Let $(G', \sigma')$ be the signed graph obtained from $(G, \sigma)$ by deleting $u$ and adding the edge $vw$ (if it was not already present).

Observe that if $vw$ was already present in $(G, \sigma)$, then it is not possible for $(G',\sigma')$ to be isomorphic to a bad $K_4$ since $v$ and $w$ have now degree 2 in $G'$. In case we do add the edge $vw$, we choose its sign in such a way we do not create any bad $K_4$. This means that if $G'$ is isomorphic to $K_4$, then we choose the sign of $vw$ so that one of the $4$-cycles of $(G', \sigma')$ becomes negative. Otherwise, we assign any sign to $vw$ in $(G', \sigma')$.

In all cases, $(G', \sigma')$ cannot be a bad $K_4$, and, hence, by minimality of $(G,\sigma)$, there exists $f: (G', \sigma') \rightarrow (SP_5^+, \square^+)$. Because $vw$ is an edge, we know that $f(v) \neq f(w)$. 
Note that this $f$ also stands as a homomorphism of 
$(G - u, \sigma)$ to $(SP_5^+, \square^+)$. 
Now, since $(SP_5^+, \square^+)$ has property $\hat{P}_{2,2}$ according to Proposition~\ref{prop property pnk}, we can extend 
$f$ to a homomorphism of $(G, \sigma)$ to $(SP_5^+, \square^+)$, a contradiction.
\end{proof}

 \begin{figure}[h!]
 	\centering
	
            
 	\subfloat[\label{subfig:c2}]{
      	\begin{tikzpicture}[inner sep=0.7mm,scale=0.8]
     		\node[draw, circle, black, fill=white, line width=1pt](u1) at (0,0)[label=above:{ $v_1$}]{};
     		\node[draw, circle, fill, black, line width=1pt](u2) at (1.5,0)[label=above:{ $v_3$}]{};
     		\node[draw, circle, fill, black, line width=1pt](u3) at (3,1)[label=above:{ $v_5$}]{};
     		\node[draw, circle, fill, black, line width=1pt](u4) at (3,-1)[label=below:{ $v_6$}]{};
     		\node[draw, circle, fill, black, line width=1pt](u5) at (4.5,0)[label=above:{ $v_4$}]{};
     		\node[draw, circle, fill, black, fill=white, line width=1pt](u6) at (6,0)[label=above:{ $v_2$}]{};
            
             \draw[-, line width=1pt,black] (u1) -- (u2); 
             \draw[-, line width=1pt,black] (u2) -- (u3);
             \draw[-, line width=1pt,black] (u2) -- (u4);
             \draw[-, line width=1pt,black] (u3) -- (u4);
             \draw[-, line width=1pt,black] (u3) -- (u5);
             \draw[-, line width=1pt,black] (u4) -- (u5);
             \draw[-, line width=1pt,black] (u5) -- (u6);
     	\end{tikzpicture}
     }
    \hspace{20pt}
 	\subfloat[\label{subfig:c3}]{
      	\begin{tikzpicture}[inner sep=0.7mm,scale=0.8]
     		\node[draw, circle, fill, black, line width=1pt](u1) at (0,0)[label=below:{ $v_4$}]{};
     		\node[draw, circle, fill, black, line width=1pt](u2) at (1.5,0)[label=below:{ $v_5$}]{};
     		\node[draw, circle, fill, black, line width=1pt](u3) at (0.75,1.5)[label=right:{ $v_6$}]{};
     		\node[draw, circle, black, fill=white, line width=1pt](u4) at (-1.5,0)[label=below:{ $v_1$}]{};
     		\node[draw, circle, black, fill=white, line width=1pt](u5) at (3,0)[label=below:{ $v_2$}]{};
     		\node[draw, circle, black, fill=white, line width=1pt](u6) at (0.75,3)[label=above:{ $v_3$}]{};
            
             \draw[-, line width=1pt,black] (u1) -- (u2); 
             \draw[-, line width=1pt,black] (u2) -- (u3);
             \draw[-, line width=1pt,black] (u3) -- (u1);
             \draw[-, line width=1pt,black] (u1) -- (u4); 
             \draw[-, line width=1pt,black] (u2) -- (u5);
             \draw[-, line width=1pt,black] (u3) -- (u6);
     	\end{tikzpicture}
    }
    \hspace{20pt}
 	\subfloat[\label{subfig:c4}]{
      	\begin{tikzpicture}[inner sep=0.7mm,scale=0.8]
     		\node[draw, circle, fill, black, line width=1pt](u1) at (0,0)[label=above:{ $v_5$}]{};
     		\node[draw, circle, fill, black, line width=1pt](u2) at (2,0)[label=above:{ $v_6$}]{};
     		\node[draw, circle, fill, black, line width=1pt](u3) at (2,-2)[label=below:{ $v_7$}]{};
     		\node[draw, circle, fill, black, line width=1pt](u4) at (0,-2)[label=below:{ $v_8$}]{};
     		\node[draw, circle, black, fill=white, line width=1pt](u5) at (-1.5,0)[label=above:{ $v_1$}]{};
     		\node[draw, circle, black, fill=white, line width=1pt](u6) at (3.5,0)[label=above:{ $v_2$}]{};
     		\node[draw, circle, black, fill=white, line width=1pt](u7) at (3.5,-2)[label=below:{ $v_3$}]{};
     		\node[draw, circle, black, fill=white, line width=1pt](u8) at (-1.5,-2)[label=below:{ $v_4$}]{};
            
             \draw[-, line width=1pt,black] (u1) -- (u2); 
             \draw[-, line width=1pt,black] (u2) -- (u3);
             \draw[-, line width=1pt,black] (u3) -- (u4);
             \draw[-, line width=1pt,black] (u4) -- (u1); 
             \draw[-, line width=1pt,black] (u1) -- (u5); 
             \draw[-, line width=1pt,black] (u2) -- (u6);
             \draw[-, line width=1pt,black] (u3) -- (u7);
             \draw[-, line width=1pt,black] (u4) -- (u8); 
     	\end{tikzpicture}
    }
    \hspace{20pt}
 	\subfloat[\label{subfig:c5}]{
      	\begin{tikzpicture}[inner sep=0.7mm,scale=0.8]
     		\node[draw, circle, black, fill=white, line width=1pt](u1) at (0,0)[label=above:{ $v_1$}]{};
     		\node[draw, circle, black, fill=white, line width=1pt](u2) at (2,0)[label=above:{ $v_2$}]{};
     		\node[draw, circle, fill, black, line width=1pt](u3) at (1,-1)[label=left:{ $v_9$}]{};
             \draw[-, line width=1pt,black] (u1) -- (u3); 
             \draw[-, line width=1pt,black] (u2) -- (u3); 
     		
     		\node[draw, circle, fill, black, line width=1pt](u5) at (1,-5)[label=left:{ $v_{12}$}]{};
     		\node[draw, circle, black, fill=white, line width=1pt](u6) at (0,-6)[label=below:{ $v_8$}]{};
     		\node[draw, circle, black, fill=white, line width=1pt](u7) at (2,-6)[label=below:{ $v_7$}]{};
             \draw[-, line width=1pt,black] (u6) -- (u5); 
             \draw[-, line width=1pt,black] (u7) -- (u5); 
             
     		\node[draw, circle, fill, black, line width=1pt](u4) at (1.5,-3)[label=left:{ $v_{14}$}]{};
             \draw[-, line width=1pt,black] (u3) -- (u4); 
             \draw[-, line width=1pt,black] (u5) -- (u4); 
     		
     		\node[draw, circle, black, fill=white, line width=1pt](u8) at (5,0)[label=above:{ $v_3$}]{};
     		\node[draw, circle, black, fill=white, line width=1pt](u9) at (7,0)[label=above:{ $v_4$}]{};
     		\node[draw, circle, fill, black, line width=1pt](u10) at (6,-1)[label=right:{ $v_{10}$}]{};
             \draw[-, line width=1pt,black] (u8) -- (u10); 
             \draw[-, line width=1pt,black] (u9) -- (u10); 
     		
     		\node[draw, circle, fill, black, line width=1pt](u12) at (6,-5)[label=right:{ $v_{11}$}]{};
     		\node[draw, circle, black, fill=white, line width=1pt](u13) at (5,-6)[label=below:{ $v_6$}]{};
     		\node[draw, circle, black, fill=white, line width=1pt](u14) at (7,-6)[label=below:{ $v_5$}]{};
             \draw[-, line width=1pt,black] (u13) -- (u12); 
             \draw[-, line width=1pt,black] (u14) -- (u12); 
     		
     		\node[draw, circle, fill, black, line width=1pt](u11) at (5.5,-3)[label=right:{ $v_{13}$}]{};
             \draw[-, line width=1pt,black] (u10) -- (u11); 
             \draw[-, line width=1pt,black] (u12) -- (u11); 
             
             \draw[-, line width=1pt,black] (u4) -- (u11);

     	\end{tikzpicture}
     }
    
     \caption{Configurations reduced for proving Theorem~\ref{th signed cubic+}. Black vertices are vertices having their whole neighborhood being part of the configuration. White vertices may have neighbors not depicted in the configuration.}
     \label{figure:configs}
 \end{figure}
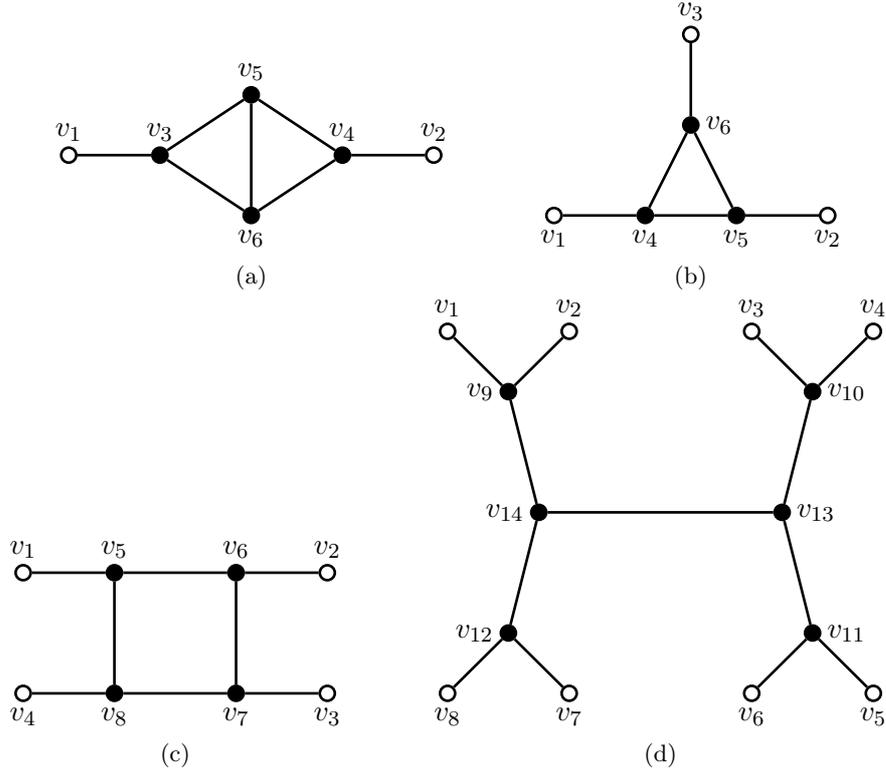

Thus, from now on we can assume that $(G, \sigma)$ is cubic.
To finish off the proof, we prove that $G$ cannot contain any of the configurations depicted in Figure~\ref{figure:configs}. 
Throughout the rest of this section, whenever dealing with one of these configurations, we do so by employing the terminology given in the figure.
It is important to emphasize that, in these configurations, white vertices are vertices that can have neighbors outside the configuration, while the whole neighborhood of the black vertices is as displayed in the configuration.
In particular, some of the white vertices could be the same vertices, or be adjacent to each other.


We proceed with the configuration depicted in Figure~\ref{figure:configs}(a).

\begin{lemma}\label{lem config2 reduced}
$(G, \sigma)$ does not contain the configuration depicted in Figure~\ref{figure:configs}(a). 
\end{lemma}

\begin{proof}
Suppose the contrary, i.e., assume that $(G, \sigma)$ 
contains the configuration depicted in Figure~\ref{figure:configs}(a). 
Let $(G', \sigma')$ be the signed graph obtained from $(G, \sigma)$ by deleting the vertices $v_3$, $v_4$, $v_5$ and $v_6$  and adding the edge $v_1v_2$ (if it was not already present).
In case $v_1v_2$ does not exist in $(G, \sigma)$,
then, in $(G', \sigma')$, just as in the proof of Lemma~\ref{lem deg 3}, 
we choose the sign of $v_1v_2$ so that $(G', \sigma')$ is not a bad $K_4$.
Thus, by minimality there exists a homomorphism
$f: (G', \sigma) \rightarrow (SP_5^+, \square^+)$.
Because $(SP_5^+, \square^+)$ is edge-transitive, 
without loss of generality we may assume  $f(v_1) = \infty$ and $f(v_2) = \overline{1}$. 
Besides, if needed, we can switch some vertices of $(G, \sigma)$ to ensure
$$\sigma(v_1v_3) = \sigma(v_3v_5) = \sigma(v_4v_5) = \sigma(v_5v_6) = +.$$
More precisely, we first switch $v_3$ if $\sigma(v_1v_3)=-$, then switch $v_5$ if $\sigma(v_3v_5)=-$, then switch $v_4$ if $\sigma(v_4v_5)=-$, and finally switch $v_6$ in case $\sigma(v_5v_6)=-$.

We first set $f(v_5) = \infty$. We now choose $\overline{i} , \overline{j}, \overline{k} \in V(SP_5^+) \setminus \{\infty\}$ so that $\overline{i}$ is a $\sigma(v_2v_4)$-neighbor of $f(v_2)=\overline{1}$, 
$\overline{j}$ is a $\sigma(v_4v_6)$-neighbor of $\overline{i}$,
and $\overline{k}$ is a $\sigma(v_3v_6)$-neighbor of $\overline{j}$. 
Now, setting $f(v_4)=\overline{i}$, $f(v_6) = \overline{j}$ and $f(v_3)=\overline{k}$,
 extends $f$ to a homomorphism of 
$(G, \sigma)$ to $(SP_5^+, \square^+)$, a contradiction. 
\end{proof}

Before proceeding with the next configuration, we first need to state a useful observation that deals with signatures $(P_3,\sigma)$ of the $3$-path $P_3=u_1u_2u_3u_4$. 

\begin{observation}\label{obs almost nice}
 Let $g$ be a partial function of $V(P_3)$ to $V(SP_5)$
 where only $u_1$ and $u_4$ get an image by $g$.
 Assume $g(u_1)= \overline{i}$ and 
 $g(u_4)= \overline{j}$ for some
 $\overline{i}, \overline{j} \in V(SP_5)$. Then,
 regardless of $\overline{i}$ and $\overline{j}$,
  it is possible to extend $g$ to an sp-homomorphism of
 $(P_3, \sigma)$ to $(SP_5, \square)$ unless 
 $\sigma(u_1u_2) = \sigma(u_2u_3) = \sigma(u_3u_4)$ and $\overline{i}=\overline{j}$. 
\end{observation}

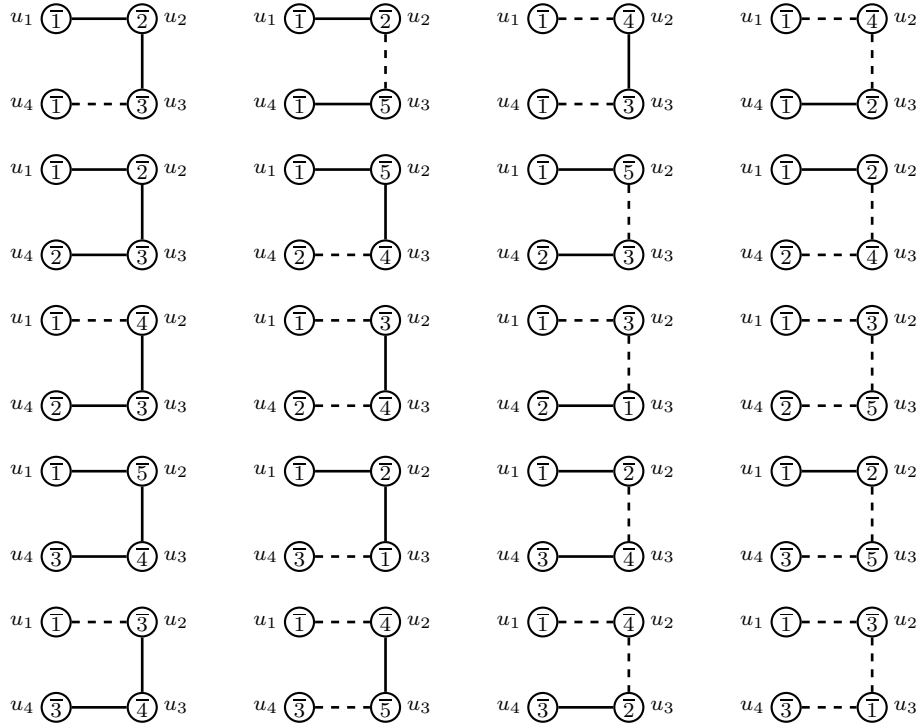
\begin{figure}[!ht]
\centering
\begin{tikzpicture}[every node/.style={draw, circle, inner sep=1pt}, thick, scale=0.8]

\node[label=left:{\scriptsize $u_1$}] (1) at (135:1) {\scriptsize $\overline{1}$};
\node[label=right:{\scriptsize $u_2$}] (2) at (45:1) {\scriptsize $\overline{2}$};
\node[label=right:{\scriptsize $u_3$}] (3) at (-45:1) {\scriptsize $\overline{3}$};
\node[label=left:{\scriptsize $u_4$}] (4) at (-135:1) {\scriptsize $\overline{1}$};
\draw[-,line width=1pt,black] (1) to (2);
\draw[-,line width=1pt,black] (2) to (3);
\draw[-,line width=1pt,black,dashed] (4) to (3);

\tikzset{xshift=4cm}
\node[label=left:{\scriptsize $u_1$}] (1) at (135:1) {\scriptsize $\overline{1}$};
\node[label=right:{\scriptsize $u_2$}] (2) at (45:1) {\scriptsize $\overline{2}$};
\node[label=right:{\scriptsize $u_3$}] (3) at (-45:1) {\scriptsize $\overline{5}$};
\node[label=left:{\scriptsize $u_4$}] (4) at (-135:1) {\scriptsize $\overline{1}$};
\draw[-,line width=1pt,black] (1) to (2);
\draw[-,line width=1pt,black,dashed] (2) to (3);
\draw[-,line width=1pt,black] (4) to (3);

\tikzset{xshift=4cm}
\node[label=left:{\scriptsize $u_1$}] (1) at (135:1) {\scriptsize $\overline{1}$};
\node[label=right:{\scriptsize $u_2$}] (2) at (45:1) {\scriptsize $\overline{4}$};
\node[label=right:{\scriptsize $u_3$}] (3) at (-45:1) {\scriptsize $\overline{3}$};
\node[label=left:{\scriptsize $u_4$}] (4) at (-135:1) {\scriptsize $\overline{1}$};
\draw[-,line width=1pt,black,dashed] (1) to (2);
\draw[-,line width=1pt,black] (2) to (3);
\draw[-,line width=1pt,black,dashed] (4) to (3);

\tikzset{xshift=4cm}
\node[label=left:{\scriptsize $u_1$}] (1) at (135:1) {\scriptsize $\overline{1}$};
\node[label=right:{\scriptsize $u_2$}] (2) at (45:1) {\scriptsize $\overline{4}$};
\node[label=right:{\scriptsize $u_3$}] (3) at (-45:1) {\scriptsize $\overline{2}$};
\node[label=left:{\scriptsize $u_4$}] (4) at (-135:1) {\scriptsize $\overline{1}$};
\draw[-,line width=1pt,black,dashed] (1) to (2);
\draw[-,line width=1pt,black,dashed] (2) to (3);
\draw[-,line width=1pt,black] (4) to (3);

\tikzset{xshift=-12cm,yshift=-2.5cm}
\node[label=left:{\scriptsize $u_1$}] (1) at (135:1) {\scriptsize $\overline{1}$};
\node[label=right:{\scriptsize $u_2$}] (2) at (45:1) {\scriptsize $\overline{2}$};
\node[label=right:{\scriptsize $u_3$}] (3) at (-45:1) {\scriptsize $\overline{3}$};
\node[label=left:{\scriptsize $u_4$}] (4) at (-135:1) {\scriptsize $\overline{2}$};
\draw[-,line width=1pt,black] (1) to (2);
\draw[-,line width=1pt,black] (2) to (3);
\draw[-,line width=1pt,black] (4) to (3);

\tikzset{xshift=4cm}
\node[label=left:{\scriptsize $u_1$}] (1) at (135:1) {\scriptsize $\overline{1}$};
\node[label=right:{\scriptsize $u_2$}] (2) at (45:1) {\scriptsize $\overline{5}$};
\node[label=right:{\scriptsize $u_3$}] (3) at (-45:1) {\scriptsize $\overline{4}$};
\node[label=left:{\scriptsize $u_4$}] (4) at (-135:1) {\scriptsize $\overline{2}$};
\draw[-,line width=1pt,black] (1) to (2);
\draw[-,line width=1pt,black] (2) to (3);
\draw[-,line width=1pt,black,dashed] (4) to (3);

\tikzset{xshift=4cm}
\node[label=left:{\scriptsize $u_1$}] (1) at (135:1) {\scriptsize $\overline{1}$};
\node[label=right:{\scriptsize $u_2$}] (2) at (45:1) {\scriptsize $\overline{5}$};
\node[label=right:{\scriptsize $u_3$}] (3) at (-45:1) {\scriptsize $\overline{3}$};
\node[label=left:{\scriptsize $u_4$}] (4) at (-135:1) {\scriptsize $\overline{2}$};
\draw[-,line width=1pt,black] (1) to (2);
\draw[-,line width=1pt,black,dashed] (2) to (3);
\draw[-,line width=1pt,black] (4) to (3);

\tikzset{xshift=4cm}
\node[label=left:{\scriptsize $u_1$}] (1) at (135:1) {\scriptsize $\overline{1}$};
\node[label=right:{\scriptsize $u_2$}] (2) at (45:1) {\scriptsize $\overline{2}$};
\node[label=right:{\scriptsize $u_3$}] (3) at (-45:1) {\scriptsize $\overline{4}$};
\node[label=left:{\scriptsize $u_4$}] (4) at (-135:1) {\scriptsize $\overline{2}$};
\draw[-,line width=1pt,black] (1) to (2);
\draw[-,line width=1pt,black,dashed] (2) to (3);
\draw[-,line width=1pt,black,dashed] (4) to (3);

\tikzset{xshift=-12cm,yshift=-2.5cm}
\node[label=left:{\scriptsize $u_1$}] (1) at (135:1) {\scriptsize $\overline{1}$};
\node[label=right:{\scriptsize $u_2$}] (2) at (45:1) {\scriptsize $\overline{4}$};
\node[label=right:{\scriptsize $u_3$}] (3) at (-45:1) {\scriptsize $\overline{3}$};
\node[label=left:{\scriptsize $u_4$}] (4) at (-135:1) {\scriptsize $\overline{2}$};
\draw[-,line width=1pt,black,dashed] (1) to (2);
\draw[-,line width=1pt,black] (2) to (3);
\draw[-,line width=1pt,black] (4) to (3);

\tikzset{xshift=4cm}
\node[label=left:{\scriptsize $u_1$}] (1) at (135:1) {\scriptsize $\overline{1}$};
\node[label=right:{\scriptsize $u_2$}] (2) at (45:1) {\scriptsize $\overline{3}$};
\node[label=right:{\scriptsize $u_3$}] (3) at (-45:1) {\scriptsize $\overline{4}$};
\node[label=left:{\scriptsize $u_4$}] (4) at (-135:1) {\scriptsize $\overline{2}$};
\draw[-,line width=1pt,black,dashed] (1) to (2);
\draw[-,line width=1pt,black] (2) to (3);
\draw[-,line width=1pt,black,dashed] (4) to (3);

\tikzset{xshift=4cm}
\node[label=left:{\scriptsize $u_1$}] (1) at (135:1) {\scriptsize $\overline{1}$};
\node[label=right:{\scriptsize $u_2$}] (2) at (45:1) {\scriptsize $\overline{3}$};
\node[label=right:{\scriptsize $u_3$}] (3) at (-45:1) {\scriptsize $\overline{1}$};
\node[label=left:{\scriptsize $u_4$}] (4) at (-135:1) {\scriptsize $\overline{2}$};
\draw[-,line width=1pt,black,dashed] (1) to (2);
\draw[-,line width=1pt,black,dashed] (2) to (3);
\draw[-,line width=1pt,black] (4) to (3);

\tikzset{xshift=4cm}
\node[label=left:{\scriptsize $u_1$}] (1) at (135:1) {\scriptsize $\overline{1}$};
\node[label=right:{\scriptsize $u_2$}] (2) at (45:1) {\scriptsize $\overline{3}$};
\node[label=right:{\scriptsize $u_3$}] (3) at (-45:1) {\scriptsize $\overline{5}$};
\node[label=left:{\scriptsize $u_4$}] (4) at (-135:1) {\scriptsize $\overline{2}$};
\draw[-,line width=1pt,black,dashed] (1) to (2);
\draw[-,line width=1pt,black,dashed] (2) to (3);
\draw[-,line width=1pt,black,dashed] (4) to (3);

\tikzset{xshift=-12cm,yshift=-2.5cm}
\node[label=left:{\scriptsize $u_1$}] (1) at (135:1) {\scriptsize $\overline{1}$};
\node[label=right:{\scriptsize $u_2$}] (2) at (45:1) {\scriptsize $\overline{5}$};
\node[label=right:{\scriptsize $u_3$}] (3) at (-45:1) {\scriptsize $\overline{4}$};
\node[label=left:{\scriptsize $u_4$}] (4) at (-135:1) {\scriptsize $\overline{3}$};
\draw[-,line width=1pt,black] (1) to (2);
\draw[-,line width=1pt,black] (2) to (3);
\draw[-,line width=1pt,black] (4) to (3);

\tikzset{xshift=4cm}
\node[label=left:{\scriptsize $u_1$}] (1) at (135:1) {\scriptsize $\overline{1}$};
\node[label=right:{\scriptsize $u_2$}] (2) at (45:1) {\scriptsize $\overline{2}$};
\node[label=right:{\scriptsize $u_3$}] (3) at (-45:1) {\scriptsize $\overline{1}$};
\node[label=left:{\scriptsize $u_4$}] (4) at (-135:1) {\scriptsize $\overline{3}$};
\draw[-,line width=1pt,black] (1) to (2);
\draw[-,line width=1pt,black] (2) to (3);
\draw[-,line width=1pt,black,dashed] (4) to (3);

\tikzset{xshift=4cm}
\node[label=left:{\scriptsize $u_1$}] (1) at (135:1) {\scriptsize $\overline{1}$};
\node[label=right:{\scriptsize $u_2$}] (2) at (45:1) {\scriptsize $\overline{2}$};
\node[label=right:{\scriptsize $u_3$}] (3) at (-45:1) {\scriptsize $\overline{4}$};
\node[label=left:{\scriptsize $u_4$}] (4) at (-135:1) {\scriptsize $\overline{3}$};
\draw[-,line width=1pt,black] (1) to (2);
\draw[-,line width=1pt,black,dashed] (2) to (3);
\draw[-,line width=1pt,black] (4) to (3);

\tikzset{xshift=4cm}
\node[label=left:{\scriptsize $u_1$}] (1) at (135:1) {\scriptsize $\overline{1}$};
\node[label=right:{\scriptsize $u_2$}] (2) at (45:1) {\scriptsize $\overline{2}$};
\node[label=right:{\scriptsize $u_3$}] (3) at (-45:1) {\scriptsize $\overline{5}$};
\node[label=left:{\scriptsize $u_4$}] (4) at (-135:1) {\scriptsize $\overline{3}$};
\draw[-,line width=1pt,black] (1) to (2);
\draw[-,line width=1pt,black,dashed] (2) to (3);
\draw[-,line width=1pt,black,dashed] (4) to (3);

\tikzset{xshift=-12cm,yshift=-2.5cm}
\node[label=left:{\scriptsize $u_1$}] (1) at (135:1) {\scriptsize $\overline{1}$};
\node[label=right:{\scriptsize $u_2$}] (2) at (45:1) {\scriptsize $\overline{3}$};
\node[label=right:{\scriptsize $u_3$}] (3) at (-45:1) {\scriptsize $\overline{4}$};
\node[label=left:{\scriptsize $u_4$}] (4) at (-135:1) {\scriptsize $\overline{3}$};
\draw[-,line width=1pt,black,dashed] (1) to (2);
\draw[-,line width=1pt,black] (2) to (3);
\draw[-,line width=1pt,black] (4) to (3);

\tikzset{xshift=4cm}
\node[label=left:{\scriptsize $u_1$}] (1) at (135:1) {\scriptsize $\overline{1}$};
\node[label=right:{\scriptsize $u_2$}] (2) at (45:1) {\scriptsize $\overline{4}$};
\node[label=right:{\scriptsize $u_3$}] (3) at (-45:1) {\scriptsize $\overline{5}$};
\node[label=left:{\scriptsize $u_4$}] (4) at (-135:1) {\scriptsize $\overline{3}$};
\draw[-,line width=1pt,black,dashed] (1) to (2);
\draw[-,line width=1pt,black] (2) to (3);
\draw[-,line width=1pt,black,dashed] (4) to (3);

\tikzset{xshift=4cm}
\node[label=left:{\scriptsize $u_1$}] (1) at (135:1) {\scriptsize $\overline{1}$};
\node[label=right:{\scriptsize $u_2$}] (2) at (45:1) {\scriptsize $\overline{4}$};
\node[label=right:{\scriptsize $u_3$}] (3) at (-45:1) {\scriptsize $\overline{2}$};
\node[label=left:{\scriptsize $u_4$}] (4) at (-135:1) {\scriptsize $\overline{3}$};
\draw[-,line width=1pt,black,dashed] (1) to (2);
\draw[-,line width=1pt,black,dashed] (2) to (3);
\draw[-,line width=1pt,black] (4) to (3);

\tikzset{xshift=4cm}
\node[label=left:{\scriptsize $u_1$}] (1) at (135:1) {\scriptsize $\overline{1}$};
\node[label=right:{\scriptsize $u_2$}] (2) at (45:1) {\scriptsize $\overline{3}$};
\node[label=right:{\scriptsize $u_3$}] (3) at (-45:1) {\scriptsize $\overline{1}$};
\node[label=left:{\scriptsize $u_4$}] (4) at (-135:1) {\scriptsize $\overline{3}$};
\draw[-,line width=1pt,black,dashed] (1) to (2);
\draw[-,line width=1pt,black,dashed] (2) to (3);
\draw[-,line width=1pt,black,dashed] (4) to (3);

\end{tikzpicture}
\caption{All cases for the proof of Observation~\ref{obs almost nice}. Solid edges are positive edges. Dashed edges are negative edges.}
\label{fig:P3}
\end{figure}

\begin{proof}
Due to the transitivity properties of $(SP_5, \square)$, it is sufficient to focus on the cases where $g(u_1)=\overline{1}$ and $g(u_4)\in\{\overline{1},\overline{2},\overline{3}\}$. Figure~\ref{fig:P3} illustrates the main cases to consider. The first row displays the cases where $g(u_4)=\overline{1}$, the second and third rows display the cases where $g(u_4)=\overline{2}$, and the fourth and fifth rows display the cases where $g(u_4)=\overline{3}$. 
\end{proof}

\begin{lemma}\label{lem config3}
$(G, \sigma)$ does not contain the configuration depicted in Figure~\ref{figure:configs}(b). 
\end{lemma}

\begin{proof}
Suppose the contrary, i.e., assume that $(G, \sigma)$ 
contains the configuration depicted in Figure~\ref{figure:configs}(b). 
Because $(G, \sigma)$ does not contain the configuration
depicted in Figure~\ref{figure:configs}(a) according to Lemma~\ref{lem config2 reduced}, note that the vertices $v_1$, $v_2$ and $v_3$ must be distinct.
Let $(G', \sigma')$ be the signed graph obtained from $(G, \sigma)$ by deleting the vertices $v_4$, $v_5$ and $v_6$ and adding the edge $v_1v_2$ (if it was not already present).
In case we do add this edge $v_1v_2$ to $(G', \sigma')$, then, as earlier, we choose its sign so that, in case $G'=K_4$, the signed graph $(G', \sigma')$ is not a bad $K_4$.
Then, by minimality, there is a homomorphism
$f: (G', \sigma') \rightarrow (SP_5^+, \square^+)$.
Since $(SP_5^+, \square^+)$ is transitive, we may assume that $f(v_3)\neq\infty$. Moreover, since $(SP_5^+ - \infty, \square^+) = (SP_5, \square)$ is sp-edge-transitive and sp-isomorphic to $(SP_5,V(SP_5)\setminus\square)$, we may assume that $f(v_1) = \overline{1}$ and $f(v_2) = \overline{2}$. Finally, we may (if needed) switch some vertices of the configuration so that 
$$\sigma(v_1v_4) = \sigma(v_4v_5) = \sigma(v_4v_6) = +.$$ 

By Observation~\ref{obs almost nice}, we can extend $f$ to a homomorphism of $(G, \sigma)$ to $(SP_5^+, \square^+)$ unless $f(v_3)=\overline{2}$ and $\sigma(v_2v_5)=\sigma(v_5v_6)=\sigma(v_3v_6)$.
This leads us to the following two cases:

\begin{enumerate}
    \item $f(v_3) = \overline{2}$ and $\sigma(v_2v_5)=\sigma(v_5v_6)=\sigma(v_3v_6) = +$. 
    
    In this case, we set $f(v_4) = \overline{2}$ in $(G', \sigma')$. The homomorphism can then be extended to $(G, \sigma)$ by setting $f(v_5) = \overline{3}$ and $f(v_6) = \infty$.

    \item $f(v_3) = \overline{2}$ and $\sigma(v_2v_5)=\sigma(v_5v_6)=\sigma(v_3v_6) = -$. 
    
    In this case, in $(G', \sigma')$ we first switch $v_4$ and $v_6$ before setting $f(v_4) = \overline{3}$. The homomorphism can then be extended to $(G, \sigma)$ by setting  $f(v_5) = \overline{5}$ and $f(v_6) = \infty$.
\end{enumerate}

In all cases, it is thus possible to extend $f$ to a homomorphism of $(G, \sigma)$ to $(SP_5^+, \square^+)$. This is a contradiction. 
\end{proof}

In order to reduce the next configuration, we need the following:

\begin{observation}\label{obs SP5 half nice}
For every two distinct $\overline{i}, \overline{j}$ in $V(SP_5)$ and $\{\alpha, \beta\} = \{+,-\}$, 
 we have  $N^{\alpha}(\overline{i}) \cap N^{\beta}(\overline{j}) \neq \emptyset $. 
\end{observation}

\begin{proof}
Due to the structure of $SP_5$, it is sufficient to prove the statement for $\overline{i}=\overline{1}$ and $\overline{j}\in\{\overline{2},\overline{3}\}$. In both cases we have $N^+(\overline{1})\cap N^-(\overline{j})=\{\overline{5}\}$ and $N^-(\overline{1})\cap N^+(\overline{j})=\{\overline{j+1}\}$
\end{proof}


\begin{lemma}\label{lem config4 reduced}
$(G, \sigma)$ does not contain the configuration depicted in Figure~\ref{figure:configs}(c). 
\end{lemma}

\begin{proof}
Suppose the contrary, i.e., assume that $(G, \sigma)$ 
contains the configuration depicted in Figure~\ref{figure:configs}(c). 
As $(G, \sigma)$ does not contain the configuration
depicted in Figure~\ref{figure:configs}(b) according to Lemma~\ref{lem config3}, the vertices $v_1$ and $v_2$ must be distinct in $(G, \sigma)$, and similarly for the vertices $v_3$ and $v_4$.
Let $(G', \sigma')$ be the signed graph obtained from $(G, \sigma)$ by deleting the vertices $v_5$, $v_6$, $v_7$ and $v_8$ and adding the edge $v_1v_2$ and $v_3v_4$ (if they were not already present).
As before, we choose the sign of each edge we add in such a way that $(G', \sigma')$ is not a bad $K_4$.
This way, by minimality, there is a homomorphism
$f: (G', \sigma') \rightarrow (SP_5^+, \square^+)$.
We know that 
$f(v_1) \neq f(v_2)$ and $f(v_3) \neq f(v_4)$. This brings us to two cases without loss of generality. 

\begin{enumerate}
    \item $f(v_3)\notin\{f(v_1), f(v_2)\}$.
    
    Without loss of generality, we may assume  $f(v_1) = \overline{1}$ and $f(v_3) = \infty$.  
    Assume that $f(v_2) = \overline{j}$ for some $\overline{j} \not \in \{\infty, \overline{1}\}$. 
    Start by switching some of $v_5,v_6,v_7,v_8$ (if needed) to make sure that $$\sigma(v_1v_5) = \sigma(v_5v_6) = \sigma(v_3v_7) = \sigma(v_5v_8) = +.$$ Set $f(v_5) = \infty$. 
    Now, choose some $\overline{i} \in N^{\sigma(v_4v_8)}(f(v_4)) \setminus \{\infty, \overline{j}\}$ and set $f(v_8) = \overline{i}$. 
    According to Observation~\ref{obs almost nice}, there is an sp-homomorphism $g$ of the signed path induced by the vertices $v_2,v_6,v_7,v_8$ such that $g(v_2) = \overline{j}$ and $g(v_8)=\overline{i}$.
    We can now extend $f$ to a homomorphism of $(G, \sigma)$ to $(SP_5^+, \square^+)$ by setting $f(v_6) = g(v_6)$ and $f(v_7) = g(v_7)$.
    
    
    
    \item $f(v_3)\in \{f(v_1), f(v_2)\}$. Up to the left/right symmetry, we may also assume that $f(v_4)\in\{f(v_1),f(v_2)\}$, i.e. $\{f(v_1),f(v_2)\} = \{f(v_3), f(v_4)\}$.
    
    We here consider two subcases:
    
    \begin{enumerate}
        \item $f(v_1) = f(v_3)$ and $f(v_2) = f(v_4)$. 
        
        Without loss of generality, assume  $f(v_1) = f(v_3) = \infty$ and $f(v_2) = f(v_4) = \overline{1}$.  
        Start by switching $v_5,v_7$ (if needed) to make sure that $$\sigma(v_1v_5) = \sigma(v_3v_7)=+.$$ 

        \begin{itemize}
            \item If the cycle $v_5v_6v_7v_8v_5$ is positive, then switch $v_6$ (if needed) to make sure that $\sigma(v_5v_6) \neq \sigma(v_5v_8)$ and $\sigma(v_6v_7) \neq \sigma(v_7v_8)$. Now choose $\overline{i} \in N^{\sigma(v_2v_6)}(\overline{1}) \setminus \{\infty\}$, $\overline{j} \in N^{\sigma(v_4v_8)}(\overline{1}) \setminus \{\infty, \overline{i}\}$ and set $f(v_6)=\overline{i}$ and $f(v_8)=\overline{j}$.
            According to Observation~\ref{obs SP5 half nice}, there is a way to extend $f$ to a homomorphism of $(G, \sigma)$ to $(SP_5^+, \square^+)$ by correctly setting $f(v_5), f(v_7)$. 
            
            \item If the cycle $v_5v_6v_7v_8v_5$ is negative, then switch some of $v_6, v_8$ (if needed)  to make sure that $\sigma(v_2v_6) = +$ and  $\sigma(v_4v_8)=-$. Up to exchanging $(v_1,v_5)$ with $(v_3,v_7)$, we may assume that $\sigma(v_5v_6) \neq \sigma(v_5v_8)$ and $\sigma(v_6v_7) = \sigma(v_7v_8)$. 
            On the one hand, if $\sigma(v_6v_7) = \sigma(v_7v_8) = +$, then set $f(v_6) = \overline{2}$, $f(v_7) = \overline{3}$ and $f(v_8) = \overline{4}$. 
            On the other hand, if $\sigma(v_6v_7) = \sigma(v_7v_8) = -$, then set $f(v_6) = \overline{2}$, $f(v_7) = \overline{5}$ and $f(v_8) = \overline{3}$. Now Observation~\ref{obs SP5 half nice} tells us that there is a way to extend $f$ to a homomorphism of $(G, \sigma)$ to $(SP_5^+, \square^+)$ by correctly setting $f(v_5)$. 
        \end{itemize}
        
        \item $f(v_1) = f(v_4)$ and $f(v_2) = f(v_3)$. 
        
        Without loss of generality, assume  $f(v_1) = f(v_4) = \infty$ and $f(v_2) = f(v_3) = \overline{1}$.  
        Start by switching some of $v_5,v_6,v_7,v_8$ (if needed) to make sure that $$\sigma(v_1v_5) = \sigma(v_2v_6) =\sigma(v_4v_8) = +$$ and $\sigma(v_3v_7) = -$.
        Set $f(v_6) =\overline{2}$ and $f(v_7)=\overline{3}$ if $\sigma(v_6v_7) = +$, and $f(v_7)=\overline{4}$ otherwise. In both cases, according to Observation~\ref{obs almost nice}, there is a way to extend $f$ to a homomorphism of $(G, \sigma)$ to $(SP_5^+, \square^+)$ by correctly setting $f(v_5)$ and $f(v_8)$.
    \end{enumerate}
\end{enumerate}

Thus, in all cases, it is possible to extend $f$ to a homomorphism of $(G, \sigma)$ to $(SP_5^+, \square^+)$. This is a contradiction. 
\end{proof}

We need two more results to deal with the last configuration in Figure~\ref{figure:configs}.
The first one deals with two particular signed graphs, $(X, \varphi)$ and $(X, \varphi')$.
Let $(X, \varphi)$ be the signed graph of order~$4$ consisting of a $3$-cycle $uvwu$ and of a vertex 
$x$ adjacent to $w$, where $\varphi$ is a signature such that $uvwu$ is a positive cycle. Let $(X, \varphi')$ be the signed graph obtained from $(X, \varphi)$ by switching the vertex $w$.

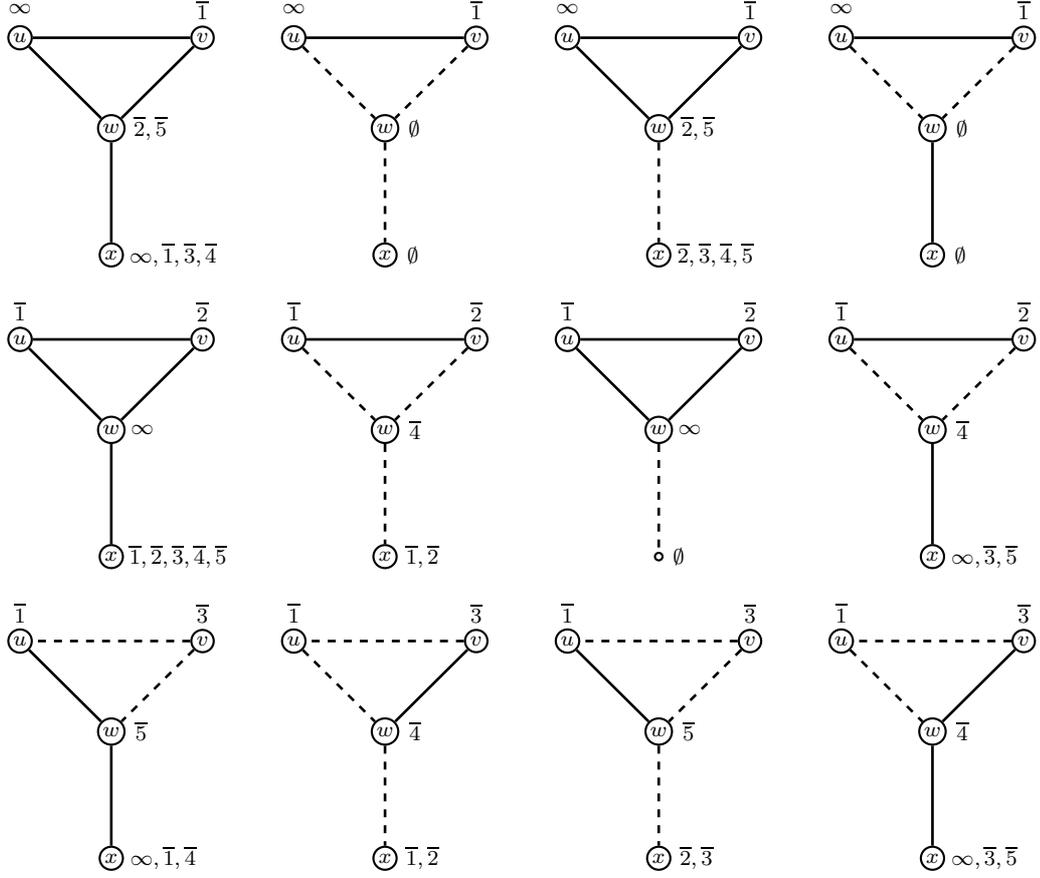
\begin{figure}[!ht]
\centering
\begin{tikzpicture}[every node/.style={draw, circle, inner sep=1pt}, thick, scale=0.8]

\node[label=above:{\scriptsize $\infty$}] (1) at (0,0) {\scriptsize $u$};
\node[label=above:{\scriptsize $\overline{1}$}] (2) at (3,0) {\scriptsize $v$};
\node[label=right:{\scriptsize $\overline{2},\overline{5}$}] (3) at (1.5,-1.5){\scriptsize $w$};
\node[label=right:{\scriptsize $\infty,\overline{1},\overline{3},\overline{4}$}] (4) at (1.5,-3.6){\scriptsize $x$};
\draw[-,line width=1pt,black] (2) to (1) to (3);
\draw[-,line width=1pt,black] (2) to (3);
\draw[-,line width=1pt,black] (3) to (4);

\tikzset{xshift=4.5cm}
\node[label=above:{\scriptsize $\infty$}] (1) at (0,0){\scriptsize $u$};
\node[label=above:{\scriptsize $\overline{1}$}] (2) at (3,0){\scriptsize $v$};
\node[label=right:{\scriptsize $\emptyset$}] (3) at (1.5,-1.5){\scriptsize $w$};
\node[label=right:{\scriptsize $\emptyset$}] (4) at (1.5,-3.6) {\scriptsize $x$};
\draw[-,line width=1pt,black] (2) to (1);
\draw[-,line width=1pt,black,dashed] (1) to (3);
\draw[-,line width=1pt,black,dashed] (2) to (3);
\draw[-,line width=1pt,black,dashed] (3) to (4);

\tikzset{xshift=4.5cm}
\node[label=above:{\scriptsize $\infty$}] (1) at (0,0){\scriptsize $u$};
\node[label=above:{\scriptsize $\overline{1}$}] (2) at (3,0){\scriptsize $v$};
\node[label=right:{\scriptsize $\overline{2},\overline{5}$}] (3) at (1.5,-1.5) {\scriptsize $w$};
\node[label=right:{\scriptsize $\overline{2},\overline{3},\overline{4},\overline{5}$}] (4) at (1.5,-3.6) {\scriptsize $x$};
\draw[-,line width=1pt,black] (2) to (1);
\draw[-,line width=1pt,black] (1) to (3);
\draw[-,line width=1pt,black] (2) to (3);
\draw[-,line width=1pt,black,dashed] (3) to (4);

\tikzset{xshift=4.5cm}
\node[label=above:{\scriptsize $\infty$}] (1) at (0,0){\scriptsize $u$};
\node[label=above:{\scriptsize $\overline{1}$}] (2) at (3,0){\scriptsize $v$};
\node[label=right:{\scriptsize $\emptyset$}] (3) at (1.5,-1.5){\scriptsize $w$};
\node[label=right:{\scriptsize $\emptyset$}] (4) at (1.5,-3.6) {\scriptsize $x$};
\draw[-,line width=1pt,black] (2) to (1);
\draw[-,line width=1pt,black,dashed] (1) to (3);
\draw[-,line width=1pt,black,dashed] (2) to (3);
\draw[-,line width=1pt,black] (3) to (4);

\tikzset{xshift=-13.5cm,yshift=-5cm}
\node[label=above:{\scriptsize $\overline{1}$}] (1) at (0,0){\scriptsize $u$};
\node[label=above:{\scriptsize $\overline{2}$}] (2) at (3,0){\scriptsize $v$};
\node[label=right:{\scriptsize $\infty$}] (3) at (1.5,-1.5) {\scriptsize $w$};
\node[label=right:{\scriptsize $\overline{1},\overline{2},\overline{3},\overline{4},\overline{5}$}] (4) at (1.5,-3.6){\scriptsize $x$};
\draw[-,line width=1pt,black] (2) to (1);
\draw[-,line width=1pt,black] (1) to (3);
\draw[-,line width=1pt,black] (2) to (3);
\draw[-,line width=1pt,black] (3) to (4);

\tikzset{xshift=4.5cm}
\node[label=above:{\scriptsize $\overline{1}$}] (1) at (0,0){\scriptsize $u$};
\node[label=above:{\scriptsize $\overline{2}$}] (2) at (3,0){\scriptsize $v$};
\node[label=right:{\scriptsize $\overline{4}$}] (3) at (1.5,-1.5){\scriptsize $w$};
\node[label=right:{\scriptsize $\overline{1},\overline{2}$}] (4) at (1.5,-3.6) {\scriptsize $x$};
\draw[-,line width=1pt,black] (2) to (1);
\draw[-,line width=1pt,black,dashed] (1) to (3);
\draw[-,line width=1pt,black,dashed] (2) to (3);
\draw[-,line width=1pt,black,dashed] (3) to (4);

\tikzset{xshift=4.5cm}
\node[label=above:{\scriptsize $\overline{1}$}] (1) at (0,0){\scriptsize $u$};
\node[label=above:{\scriptsize $\overline{2}$}] (2) at (3,0){\scriptsize $v$};
\node[label=right:{\scriptsize $\infty$}] (3) at (1.5,-1.5) {\scriptsize $w$};
\node[label=right:{\scriptsize $\emptyset$}] (4) at (1.5,-3.6) {};
\draw[-,line width=1pt,black] (2) to (1);
\draw[-,line width=1pt,black] (1) to (3);
\draw[-,line width=1pt,black] (2) to (3);
\draw[-,line width=1pt,black,dashed] (3) to (4);

\tikzset{xshift=4.5cm}
\node[label=above:{\scriptsize $\overline{1}$}] (1) at (0,0){\scriptsize $u$};
\node[label=above:{\scriptsize $\overline{2}$}] (2) at (3,0){\scriptsize $v$};
\node[label=right:{\scriptsize $\overline{4}$}] (3) at (1.5,-1.5){\scriptsize $w$};
\node[label=right:{\scriptsize $\infty,\overline{3},\overline{5}$}] (4) at (1.5,-3.6){\scriptsize $x$};
\draw[-,line width=1pt,black] (2) to (1);
\draw[-,line width=1pt,black,dashed] (1) to (3);
\draw[-,line width=1pt,black,dashed] (2) to (3);
\draw[-,line width=1pt,black] (3) to (4);

\tikzset{xshift=-13.5cm,yshift=-5cm}
\node[label=above:{\scriptsize $\overline{1}$}] (1) at (0,0){\scriptsize $u$};
\node[label=above:{\scriptsize $\overline{3}$}] (2) at (3,0){\scriptsize $v$};
\node[label=right:{\scriptsize $\overline{5}$}] (3) at (1.5,-1.5){\scriptsize $w$};
\node[label=right:{\scriptsize $\infty,\overline{1},\overline{4}$}] (4) at (1.5,-3.6) {\scriptsize $x$};
\draw[-,line width=1pt,black,dashed] (2) to (1);
\draw[-,line width=1pt,black] (1) to (3);
\draw[-,line width=1pt,black,dashed] (2) to (3);
\draw[-,line width=1pt,black] (3) to (4);

\tikzset{xshift=4.5cm}
\node[label=above:{\scriptsize $\overline{1}$}] (1) at (0,0){\scriptsize $u$};
\node[label=above:{\scriptsize $\overline{3}$}] (2) at (3,0){\scriptsize $v$};
\node[label=right:{\scriptsize $\overline{4}$}] (3) at (1.5,-1.5) {\scriptsize $w$};
\node[label=right:{\scriptsize $\overline{1},\overline{2}$}] (4) at (1.5,-3.6) {\scriptsize $x$};
\draw[-,line width=1pt,black,dashed] (2) to (1);
\draw[-,line width=1pt,black,dashed] (1) to (3);
\draw[-,line width=1pt,black] (2) to (3);
\draw[-,line width=1pt,black,dashed] (3) to (4);

\tikzset{xshift=4.5cm}
\node[label=above:{\scriptsize $\overline{1}$}] (1) at (0,0){\scriptsize $u$};
\node[label=above:{\scriptsize $\overline{3}$}] (2) at (3,0){\scriptsize $v$};
\node[label=right:{\scriptsize $\overline{5}$}] (3) at (1.5,-1.5) {\scriptsize $w$};
\node[label=right:{\scriptsize $\overline{2},\overline{3}$}] (4) at (1.5,-3.6){\scriptsize $x$};
\draw[-,line width=1pt,black,dashed] (2) to (1);
\draw[-,line width=1pt,black] (1) to (3);
\draw[-,line width=1pt,black,dashed] (2) to (3);
\draw[-,line width=1pt,black,dashed] (3) to (4);

\tikzset{xshift=4.5cm}
\node[label=above:{\scriptsize $\overline{1}$}] (1) at (0,0){\scriptsize $u$};
\node[label=above:{\scriptsize $\overline{3}$}] (2) at (3,0){\scriptsize $v$};
\node[label=right:{\scriptsize $\overline{4}$}] (3) at (1.5,-1.5){\scriptsize $w$};
\node[label=right:{\scriptsize $\infty,\overline{3},\overline{5}$}] (4) at (1.5,-3.6){\scriptsize $x$};
\draw[-,line width=1pt,black,dashed] (2) to (1);
\draw[-,line width=1pt,black,dashed] (1) to (3);
\draw[-,line width=1pt,black] (2) to (3);
\draw[-,line width=1pt,black] (3) to (4);
    
\end{tikzpicture}
\caption{All cases for the proof of Observation~\ref{obs long}. Solid edges are positive edges. Dashed edges are negative edges.}
\label{figure:long-obs}
\end{figure}

\begin{observation}\label{obs long}
Let $g$ be a partial sp-homomorphism from $(X, \varphi)$ to 
$(SP_5^+, \square^+)$, where only $u$ and $v$ have an image under $g$.
Then, up to switching the vertex $w$, it is possible to extend $g$ to an sp-homomorphism from $(X, \varphi)$ to $(SP_5^+, \square^+)$ satisfying the following: 
\begin{enumerate}[(a)]
    \item if $\{g(u), g(v)\} = \{\infty, \overline{i}\}$ and $\varphi(wx)=\varphi(uw)$, then $g(x) \notin \{\overline{i-1},\overline{i+1}\}$;
    
    \item if $\{g(u), g(v)\} = \{\infty, \overline{i}\}$ and $\varphi(wx)\neq \varphi(uw)$, then $g(x) \notin \{\infty,\overline{i}\}$;

    \item if $\{g(u), g(v)\} = \{\overline{i}, \overline{i+1}\}$ and $\varphi(wx)=\varphi(uw)$, 
    then $g(x)\neq\infty$;
    
    \item if $\{g(u), g(v)\} = \{\overline{i}, \overline{i+1}\}$ and $\varphi(wx)\neq \varphi(uw)$, 
    then $g(x) \notin \{\overline{i}, \overline{i+1},\overline{i+3}\}$;
    
    \item if $\{g(u), g(v)\} = \{\overline{i}, \overline{i+2}\}$ and $\varphi(wx)=\varphi(uw)$, 
    then $g(x) \notin \{\overline{i+2},\overline{i+4}\}$;
    
    \item if $\{g(u), g(v)\} = \{\overline{i}, \overline{i+2}\}$ and $\varphi(wx)\neq \varphi(uw)$, 
    then $g(x) \notin \{\overline{i},\overline{i-2}\}$.
 \end{enumerate}
 \end{observation}
 
\begin{proof}
Due to the symmetric structure of $(SP_5^+, \square^+)$, it is sufficient to consider the cases depicted in Figure~\ref{figure:long-obs}. For each considered value of $\{g(u),g(v)\}$ and $\varphi(wx), \varphi(uw)$, we give two signatures on $X$ ($\varphi$ and $\varphi'$) and the corresponding possible values of $g(w)$ and $g(x)$.
\end{proof}

The second result we need deals with two additional signed graphs, $(Y, \varphi)$ and $(Y, \varphi')$, obtained from $(X, \varphi)$ and $(X, \varphi')$, respectively, by adding a new vertex $y$ adjacent to $x$ through a positive edge. 

\begin{observation}\label{obs short}
Let $g$ be a partial sp-homomorphism of $(Y, \varphi)$ to 
$(SP_5^+, \square^+)$, where only $u$, $v$ and $y$ have an image under $g$.
Then, it is possible to extend $g$ to an sp-homomorphism of $(Y, \varphi)$ or $(Y, \varphi')$ to $(SP_5^+, \square^+)$.
\end{observation}

\begin{proof}
Observe that, for $\infty$ and any other vertex in $(SP_5^+, \square^+)$, the union of their positive neighborhoods is $V(SP_5^+)$. Moreover, note that, in $(SP_5^+, \square^+)$, we have 
$$\forall i\in V(SP_5), \quad N^{+}(\overline{i}) \cup N^{+}(\overline{i+1}) \cup N^{+}(\overline{i+2}) = V(SP_5^+).$$
The result now follows from Observation~\ref{obs long}. 
\end{proof}

We are now ready to reduce the final configuration. 

\begin{lemma}\label{lem config5}
$(G, \sigma)$ does not contain the configuration depicted in Figure~\ref{figure:configs}(d). 
\end{lemma}

\begin{proof}
Suppose the contrary, i.e., assume that $(G, \sigma)$ 
contains the configuration depicted in Figure~\ref{figure:configs}(d). Let $(G'', \sigma'')$ be the signed graph obtained from $(G, \sigma)$  by  adding the edges $v_1v_2$, $v_3v_4$, $v_5v_6$ and $v_7v_8$. Note that these edges were not already present in $(G, \sigma)$, since $(G, \sigma)$ cannot have $3$-cycles according to Lemma~\ref{lem config3}.
Also let $(G', \sigma')$ be the signed graph obtained from 
$(G'', \sigma'')$ by deleting the vertices $v_9$, $v_{10}$, $v_{11}$, $v_{12}$, $v_{13}$ and $v_{14}$. Observe that if a connected component of $G'$ is isomorphic to $K_4$, then $G$ must have a cut-vertex, which is not possible by Lemma~\ref{lemma:connected}.
Thus, we can freely choose the signs of the edges we have just added, without caring of whether a bad $K_4$ is created. 
Precisely, we assign signs to $v_1v_2$, $v_3v_4$, $v_5v_6$, $v_7v_8$ in such a way that the $3$-cycle $v_1v_2v_9v_1$ is negative, while the $3$-cycles $v_3v_4v_{10}v_3$, $v_5v_6v_{11}v_5$ and $v_7v_8v_{12}v_7$ are positive. 

By minimality of $(G, \sigma)$,  there is a homomorphism
$f: (G', \sigma') \rightarrow (SP_5^+, \square^+)$.
Because $(SP_5^+, \square^+)$ is edge-transitive, 
without loss of generality we may assume  $f(v_1) = \overline{1}$ and $f(v_2) = \overline{3}$. 
Note also that because the cycle $v_1v_2v_9v_1$ is negative and $v_1v_2$ is a negative edge (due to the images of $v_1,v_2$ in $(SP_5^+, \sigma)$), we can, if needed, switch $v_9$ to ensure 
$\sigma(v_1v_9)=\sigma(v_2v_9)=+$. We can also switch $v_{14}$ and/or $v_{13}$ (if needed) to ensure 
$\sigma(v_9v_{14})=\sigma(v_{13}v_{14}) = +$. 

Note that the signed subgraphs induced by $\{v_3,v_4,v_{10},v_{13}\}$ and $\{v_5,v_6,v_{11},v_{13}\}$ are exactly the signed graphs
$(X, \varphi)$ or $(X, \varphi')$ described in Observation~\ref{obs long}. If one of them does not fall into Observation~\ref{obs long}(d), then
at most five values are forbidden at $f(v_{13})$. Thus it is possible to extend $f$ to $\{v_{10},v_{11},v_{13}\}$ a homomorphism of $(G, \sigma)$ to $(SP_5^+, \square^+)$.  

Otherwise, both of them satisfy the requirements of  Observation~\ref{obs long}(d), and we can also extend $f$ to $\{v_{10},v_{11},v_{13}\}$ by setting (for example) $f(v_{13})=\infty$.

Now, observe that $\{v_7,v_8,v_{12},v_{14},v_{13}\}$ induces the graph $Y$. By Observation~\ref{obs short}, we can extend $f$ to $\{v_{12},v_{14}\}$ regardless of the value of the values of $f(v_7),f(v_8)$ and $f(v_{13})$. 

Finally, we extend $f$ to a homomorphism of $(G, \sigma)$ to $(SP_5^+, \square^+)$ by assigning $f(v_9)=\infty$ if $f(v_{14})\neq\infty$ and $f(v_9)=\overline{2}$ otherwise, which is a contradiction. 
\end{proof}

The proof of Theorem~\ref{th signed cubic+} now follows from the fact that every subcubic graph different from $K_4$ must have minimum degree~$1$ or~$2$, or must contain one of the configurations depicted in Figure~\ref{figure:configs}. Then, the previous lemmas imply that $(G, \sigma)$ cannot exist, a contradiction.




\section{Conclusion}\label{sec conclusion}
In this work, we have investigated the signed chromatic number of particular classes of graphs, namely planar graphs, triangle-free planar graphs, $K_n$-minor-free graphs, and graphs with bounded maximum degree. We have mainly considered general bounds (Theorems~\ref{th Kn-minor-free} and~\ref{th signed cubic+}) for some of these classes, and the uniqueness of bounds (Theorems~\ref{th 2ec planar bound} and~\ref{th triangle-free planar bound}) for the others. While some of our results are original ones, other ones extend known results from the literature.

Most of our results yield interesting research perspectives for the future, either because they are not tight yet, or because they lead to interesting side questions. In particular, we wonder how the bounds in Theorems~\ref{th Kn-minor-free} and~\ref{th delta-toka} should be sharpened. Regarding Theorem~\ref{th signed cubic+}, it would be interesting to determine whether $(SP_5^+, \square^+)$ is the only bound of order~$6$ for subcubic graphs. 
Regarding Theorems~\ref{th 2ec planar bound} 
and~\ref{th triangle-free planar bound}, it would be, more generally speaking, of prime importance to understand better the signed chromatic number of planar graphs, for which the currently best known lower and upper bounds are rather distant. An interesting more general question as well, could be to consider how the signed chromatic of a planar graph relates to its girth. That is, studying $\chi_s(\mathcal{P}_g)$ for any $g \geq 3$.

\section*{Acknowledgement}

The authors were partly supported by ANR project HOSIGRA (ANR-17-CE40-0022), by IFCAM project ``Applications of graph homomorphisms'' (MA/IFCAM/18/39) and by the MUNI Award in Science and Humanities of the Grant Agency of Masaryk university.

\bibliographystyle{abbrv}
\bibliography{POreferences_v2}

\end{document}